\documentclass[prd,eqsecnum,showpacs,amsmath,amsfonts,amssymb,twocolumn]{revtex4}
\usepackage[final]{graphicx}
\usepackage{hyperref}
\usepackage{bm}
\allowdisplaybreaks[1]

\newcommand{\conj}[1]{\bar{#1}}
\newcommand{\Dirac}{\delta}
\newcommand{\Heaviside}{\theta}
\newcommand{\approaches}{\rightarrow}
\renewcommand{\d}{\mathrm{d}}
\newcommand{\odiff}[2]{\frac{\d#1}{\d#2}}
\newcommand{\diff}[2]{\frac{\partial#1}{\partial#2}}

\newcommand{\programname}[1]{\texttt{\textsc{#1}}}
\newcommand{\routinename}[1]{\texttt{#1}}
\newcommand{\abs}[1]{\left\lvert{#1}\right\rvert}
\newcommand{\Journalname}[1]{\textit{#1}}

\newcommand{\jump}[1]{\left[#1\right]}
\newcommand{\const}{\mathrm{const}}
\newcommand{\orderof}[1]{O(#1)}
\newcommand{\real}{\operatorname{Re}}
\newcommand{\imag}{\operatorname{Im}}
\newcommand{\define}{:=}

\newcommand{\pdef}{\xi}
\newcommand{\mdef}{\bar \xi}

\begin{document}
\title{Scalar self-force on eccentric geodesics in Schwarzschild
spacetime: A time-domain computation}
\author{Roland \surname{Haas}}
\affiliation{Department of Physics, University of Guelph, Guelph,
Ontario, Canada N1G 2W1}
\date{April 3, 2007}
\pacs{04.25.-g, 04.40.-b, 41.60.-m, 45.50.-j, 02.60.Cb, 02.70.Bf}

\begin{abstract}
    We calculate the self-force acting on a  particle with scalar charge
    moving on a generic geodesic around a Schwarzschild black hole. This
    calculation requires an accurate computation of the retarded scalar field
    produced by the moving charge; this is done numerically with the help of a
    fourth-order convergent finite-difference scheme formulated in the time
    domain. The calculation also requires a regularization procedure, because
    the retarded field is singular on the particle's world line; this is
    handled mode-by-mode via the mode-sum regularization scheme first
    introduced by Barack and Ori. This paper presents the numerical method,
    various numerical tests, and a sample of results for mildly
    eccentric orbits as well as ``zoom-whirl'' orbits.
\end{abstract}

\maketitle

\section{Introduction}
The inspiral and capture of solar-mass compact objects by
supermassive black holes is one of the most promising and interesting sources
of gravitational
radiation to be detected by the future space-based gravitational-wave antenna
LISA~\cite{LISA-project}. For these extreme mass-ratio inspirals, one can treat
the compact object as a point mass and describe its influence on the spacetime
perturbatively. 
Going beyond the test mass limit, its motion is no longer along a geodesic of
the unperturbed spacetime of the central black hole; it is a geodesic
of the perturbed spacetime created by the presence of the moving body. 
When viewed from the unperturbed
spacetime, the small body is said to move under the influence of its
gravitational self-force.  
The self-force induces radiative losses of energy and angular momentum, 
which will eventually drive the object into the black hole.
To describe the motion of the body, including its
inspiral toward the black hole, we seek to evaluate the self-force and
calculate its effect on the motion. 
One way of doing this uses the mode-sum
regularization procedure introduced by Barack and Ori~\cite{barack:00}.
(For a comprehensive introduction of the problem, see the special issue of
\Journalname{Classical and Quantum Gravity}~\cite{CQG-special-issue}.)

In this paper, in an effort to build expertise to calculate the gravitational
self-force, we retreat to the technically simpler problem of a point
particle of mass $m$ endowed with a scalar charge $q$ orbiting a
Schwarzschild black hole of mass $M$. Following up on a previous
paper~\cite{haas:06}, we implement the numerical part of the regularization
procedure for generic orbits with a time-domain integration of the scalar-wave
equation.

\subsection{The problem}\label{sec:the-problem}
Our goal is to calculate the regularized self-force acting on a scalar point
charge in orbit around a Schwarzschild black hole. In analogy with the
gravitational case, where in a first-order (in $m/M$) perturbative calculation
the particle moves on a geodesic of the background spacetime, we take the
orbit of the particle to be a geodesic and calculate the self-force as a vector
field on this geodesic.  We start by writing the Schwarzschild
metric using the tortoise coordinate
$r^* = r + 2M\ln\left(\frac{r}{2M}-1\right)$ as
\begin{align}
    \d s^2 &= f\,\left( -\d t^2 + {\d r^*}^2 \right) +
    r^2\d\Omega^2\text{,}
\end{align}
where $f = \left( 1-\frac{2M}{r} \right)$, $\d\Omega^2 = \left( \d \theta^2 +
\sin^2\theta \d\phi^2 \right)$ is the metric on a two-sphere, and $t$, $r$, 
$\theta$ and $\phi$ are the usual Schwarzschild coordinates.
Our task is to solve the scalar wave equation
\begin{align}
    g^{\alpha\beta} \nabla_\alpha \nabla_\beta \Phi(x) &= -4\pi \mu(x)
    \text{,}\label{eqn:covariant-wave-eqn} \\
    \mu(x) &= q \int_\gamma \Dirac_4(x,z(\tau)) \d\tau\text{,}
\end{align}
where $\nabla_\alpha$ is the covariant derivative compatible with the metric
$g_{\alpha\beta}$, $\Phi(x)$ is the scalar field created by a scalar charge
$q$ which moves along a world line $\gamma: \tau \mapsto z(\tau)$ parametrized by 
proper time
$\tau$. The source term $\mu(x)$ appearing on the right-hand side
is written in terms of a scalarized four-dimensional Dirac $\Dirac$-function
$\Dirac_4(x,x') \define \Dirac(x_0-x'_0) \Dirac(x_1-x'_1) \Dirac(x_2-x'_2)
\Dirac(x_3-x'_3)/\sqrt{-\det(g_{\alpha\beta})}$.
Because of the singularity in the source term, the retarded solution to
Eq.~\eqref{eqn:covariant-wave-eqn} is singular on the world line, and the
na\"\i{}ve expression for the self-force,
\begin{align}
    F_\alpha(\tau) &= q \nabla_\alpha \Phi\bm(z(\tau)\bm)\text{,}
\end{align}
must be regularized. Following DeWitt and Brehme~\cite{dewitt:60}, Mino,
Sasaki, Tanaka~\cite{mino:97}, Quinn
and Wald~\cite{quinn:97}, Quinn~\cite{quinn:00} carried out this
regularization for the electromagnetic, scalar and gravitational radiation
reaction. In later work, Detweiler and Whiting~\cite{detweiler:03}
introduced a very useful
decomposition of the retarded solution of Eq.~\eqref{eqn:covariant-wave-eqn} in
terms of a singular part $\Phi^S$ and a regular remainder $\Phi^R$:
\begin{align}
\Phi &= \Phi^S + \Phi^R \text{.}\label{eqn:DW-decomp}
\end{align}
$\Phi^R$ is regular and differentiable at the position of the particle,
satisfies the homogeneous wave equation associated with 
Eq.~\eqref{eqn:covariant-wave-eqn}, and is solely responsible for the
self-force acting on the particle. $\Phi^S$, on the other hand, satisfies
Eq.~\eqref{eqn:covariant-wave-eqn}, is just as singular at the particle's
position as the retarded solution, and produces no force on the particle.
Rearranging Eq.~\eqref{eqn:DW-decomp} and
differentiating once, we can write the regularized self-force as
\begin{align}
F_\alpha &\define q \nabla_\alpha \Phi^R = q \bigl(\nabla_\alpha \Phi - \nabla_\alpha
    \Phi^S\bigr) \text{.}\label{eqn:regularized-self-force}
\end{align}

In a previous paper~\cite{haas:06}, we described our implementation of the
regularization procedure to find a mode-sum representation of $\nabla_\alpha
\Phi^S$ along a generic
geodesic of the Schwarzschild spacetime. Schematically, we
introduce a tetrad $e^\alpha_{(\mu)}$ and decompose the tetrad components
$\Phi_{(\mu)} \define e^\alpha_{(\mu)} \nabla_\alpha \Phi$ of the field
gradient in terms of
ordinary scalar spherical harmonics $Y_{\ell m}$:
\begin{align}
\Phi_{(\mu)}(t, r, \theta, \phi) = \sum_{\ell, m} \Phi^{\ell m}_{(\mu)}(t, r) 
    Y_{\ell m}(\theta, \phi)\text{.}
\end{align}
Each mode
$\Phi^{\ell m}_{(\mu)}(t, r)$ is finite at the position of the particle, but their sum
diverges on the world line. In~\cite{haas:06}, we derive analytic expressions
for the
mode-sum decomposition of $\Phi^S_{(\mu)}$,
\begin{align}
    \Phi^S_{(\mu)} = &q \sum_\ell \Phi^S_{(\mu),\ell} \\
    \Phi^S_{(\mu),\ell} &= A_{(\mu)} \Bigl(\ell + \frac 12\Bigr) + B_{(\mu)} +
\frac{C_{(\mu)}}{\ell + \frac 12}
\nonumber\\\mbox{}&+ \frac{D_{(\mu)}}{(\ell - \frac 12) (\ell +
\frac 32)} + \cdots\text{,}\label{eqn:singular-field}
\end{align}
where the coefficients $A_{(\mu)}$, $B_{(\mu)}$, $C_{(\mu)}$, and $D_{(\mu)}$
are independent of $\ell$; they are listed in
Appendix~\ref{sec:regularization-parameters} for convenience. 

As each mode
of $\Phi$ is finite, it is straightforward to compute the modes of the
retarded solution using numerical methods, and we will describe how this was
done in Sec.~\ref{sec:implementation}.
We use the numerical solutions in
Eq.~\eqref{eqn:regularized-self-force} to calculate the regularized
self-force, regularizing mode-by-mode:
\begin{align}
    \Phi^R_{(\mu)} &= \sum_\ell \bigl( \Phi_{(\mu), \ell} -
    \Phi^S_{(\mu), \ell} \bigr)\text{,}\label{eqn:mode-by-mode-regularization}
\end{align}
where $\Phi_{(\mu), \ell} \define \sum_m \Phi^{\ell m}_{(\mu)}
Y_{\ell m}$ (no summation over $\ell$ implied).

For numerical purposes it is convenient to define  
$\psi_{\ell m}$ by 
\begin{align}
    \Phi(x) &= \sum_{\ell=0}^{\infty} \sum_{m=-\ell}^{\ell} \frac1r \psi_{\ell m}
    Y^{\ell m}\text{,}\label{eqn:spherical-harmonics-decomposition}
\end{align}
where $Y_{\ell m}$ are the usual scalar spherical harmonics. 
After substituting in Eq.~\eqref{eqn:covariant-wave-eqn}, this
yields a reduced wave equation for the multipole
moments $\psi_{\ell m}$:
\begin{align}
    -\partial_t^2 \psi_{\ell m} &+ \partial_{r^*}^2 \psi_{\ell m} 
    - V_\ell \psi_{\ell m} =
        \nonumber \\
    & -4\pi q \frac{f_0}{r_0 E} \conj{Y}_{\ell m}(\pi/2, \phi_0) 
    \Dirac(r^* - r_0^*)\text{,}\label{eqn:reduced-wave-eqn}
\end{align}
where
\begin{align}
    V_\ell = f\left(\frac{2M}{r^3}+\frac{\ell\,(\ell+1)}{r^2}\right)
\text{.}
\end{align}
An overbar denotes complex
conjugation, $E = -u_t$ is the particle's conserved energy per unit mass,
and $u^\alpha = \odiff{z^\alpha}{\tau}$ is its four velocity.
Quantities bearing a subscript ``$0$'' are evaluated at the
particle's position; they are functions of $\tau$ that are obtained by solving
the geodesic equation 
\begin{align}
    u^\beta \nabla_\beta u^\alpha &= 0
\end{align}
in the background spacetime.
Without loss of generality, we have confined the motion of the
particle to the equatorial plane $\theta = \frac\pi2$. 

Once we have numerically solved Eq.~\eqref{eqn:reduced-wave-eqn}, we extract
numerical estimates for $\psi_{\ell m}$, $\partial_t \psi_{\ell m}$ and
$\partial_{r^*} \psi_{\ell m}$, which can then be used to find 
$\Phi_{\ell m}$, $\partial_t \Phi_{\ell m}$ and
$\partial_r \Phi_{\ell m}$. These---together with the translation table
displayed in 
Eqs.~(1.23)--(1.26) of~\cite{haas:06}, reproduced in
Appendix~\ref{sec:translation-tables}---allow us to find the tetrad components
$\Phi_{(\mu) \ell m}$ with respect to the tetrad defined by Eqs.~(1.18)--(1.21)
of~\cite{haas:06}. Eventually we regularize the \emph{multipole
coefficients}
\begin{align}
    \Phi_{(\mu) \ell} &= \sum_{m=-\ell}^{\ell} \Phi_{(\mu) \ell m}(t_0, r_0)
    Y^{\ell m}(\pi/2, \phi_0)\label{eqn:multipole-coeffs}
\end{align}
using Eq.~\eqref{eqn:mode-by-mode-regularization}; this involves the
regularization parameters listed in Eqs.~(1.30)--(1.45)
of~\cite{haas:06}, which are reproduced in
Appendix~\ref{sec:regularization-parameters}.

\subsection{Organization of this paper}
In Sec.~\ref{sec:numerical-method} we introduce the main ideas behind the
discretization scheme used in the numerical simulation.
Sec.~\ref{sec:initial-and-boundary-condition} describes the choices we make
in order to handle the problems of specifying initial data and proper boundary
conditions. The next section---Sec.~\ref{sec:implementation}---provides
details on the concrete implementation of the ideas put forth in
Secs.~\ref{sec:numerical-method}
and~\ref{sec:initial-and-boundary-condition}. In
Sec.~\ref{sec:numerical-tests} we describe the tests we performed in order
to validate our implementation of the numerical method.
Sec.~\ref{sec:sample-results} finally presents sample results for a small
number of representative simulations. 

\subsection{Future work}\label{sec:future-work}
This work, which deals with a scalar charge moving in the Schwarzschild
spacetime, is not intended to produce physically or astrophysically
interesting results. Instead, its goal is to help us evaluate the merits of
several strategies that could be used to tackle the more interesting (and
difficult) problems of electromagnetism and gravity. 

One future project we are currently exploring is to apply the formalism
developed so far to the electromagnetic self-force acting on an electric
charge. Beyond the technical complication of having to deal with a vector
field instead of a
single scalar quantity, we are also faced with the reality of having to
impose a gauge (in our case: the Lorenz gauge) and to eliminate (or at least
control) gauge violations in the numerical simulation. The first step, namely,
the calculation of the regularization parameters $A_{(\mu)}$, $B_{(\mu)}$,
$C_{(\mu)}$, and $D_{(\mu)}$ for the self-force, is currently underway. Also
underway is the calculation of the regularization parameters for he
gravitational self-force. 

Another project is the implementation of a scheme to use the
calculated self-force to update the orbital parameters of a particle on its
inspiral toward the black hole. The standard proposed approach to this problem in the
past has been to calculate the self-force on a set of geodesics which are
momentarily tangent to the particle's trajectory. The self-force calculated
in this way is then used to update the orbital elements. This ``after the fact''
calculation of the motion requires one to build (in advance) a large database of
self-force values for the anticipated set of orbital parameters that the
particle's trajectory will assume during its inspiral. Alternatively, and
conceptually more simply,  the
self-force could be calculated self-consistently along the real, accelerated
trajectory. Such an approach requires changes in the expressions of the
regularization parameters, which so far have been derived only for geodesic
orbits. We are currently investigating the merits of such an approach.

\section{Numerical method}\label{sec:numerical-method}
In this section we describe the algorithm used to integrate the reduced wave
equation [Eq.~\eqref{eqn:reduced-wave-eqn}] numerically. For the most part we
use the fourth-order algorithm introduced by Lousto~\cite{lousto:05}, with
some modifications to suit our needs. 
We choose to implement a fourth-order convergent code because second-order
convergence
for the potential $\Phi$, while much easier to achieve,  would guarantee
only first-order convergence for 
$\nabla_\alpha \Phi$, the quantity in which we are ultimately interested.
With a fourth-order convergent code we can expect to achieve third-order
convergence for $\nabla_\alpha \Phi$, which is required for an accurate
estimation of the self-force.
Numerical experiments, however, show that in practice we \emph{do} achieve
fourth-order convergence for the derivatives of $\Phi$, a fortunate
outcome that we exploit but cannot explain.

From now on, we will suppress the subscripts $\ell$ and $m$ on $V_\ell$ and
$\psi_{\ell m}$ for convenience of notation.
The wave equation consists of three parts: the wave-operator term
$(\partial_{r^*}^2 - \partial_t^2)\psi$ and the potential term $V
\psi$ on the left-hand side, and the source term on the right-hand 
side of the equation.  Of these,
the wave operator turns out to be easiest to handle, and
the source term does not create a substantial difficulty.
The term involving the potential $V$ turns out to be the
most difficult one to handle. 

Following Lousto we introduce a staggered grid with step sizes $\Delta t = \frac12 \Delta
r^* \equiv h$, which follows the characteristic lines of the wave
operator in Schwarzschild spacetime; see Fig.~\ref{fig:simpson-points} for a sketch of
a typical grid cell. 
The basic idea behind the method is
to integrate the wave equation over a unit cell of the grid, which nicely
deals with the
Dirac-$\Dirac$ source term on the right-hand side. To this end, we introduce
the Eddington-Finkelstein null coordinates $v = t + r^*$ and $u = t - r^*$ and
use them as integration variables.

\subsection{Differential operator}
Rewriting the wave operator in terms of $u$ and $v$, we find $-\partial_t^2 +
\partial_{r^*}^2 = -4 \partial_u\partial_v$, which allows us to evaluate the
integral involving the wave operator exactly. We find
\begin{align}
    \int\!\!\int_{\text{cell}} -4\partial_u\partial_v \psi \,
    \d u \, \d v =& -4[\psi(t+h, r^*) + \psi(t-h, r^*)  
    \nonumber \\
    & \mbox{} - \psi(t, r^*-h) - \psi(t, r^*+h)]
    \text{.}\label{eqn:differential-operator}
\end{align}

\subsection{Source term}
If we integrate over a cell traversed by the particle, then the source term
on the right-hand side of the equation will have a non-zero contribution.
Writing the source term as $G(t, r^*) \delta\bm(r^* - r_0^*(t)\bm)$
with
\begin{align}
    G(t,r^*) = -4\pi q \frac{f}{E r}\conj{Y}_{\ell
    m}(\pi/2,\phi_0)\text{,}
\end{align}
we find
\begin{align}
    \int\!\!\int_{\text{cell}} G \Dirac\bm(r^*-r^*_0(t)\bm) \, \d u \, \d v = &
    -\frac{8\pi q}{E} \int_{t_1}^{t_2} \frac{f_0(t)}{r_0(t)} 
    \nonumber\\\mbox{}&\times\conj{Y}_{\ell m}\bm(\pi/2, \phi_0(t)\bm) \,\d t
    \text{,}
\end{align}
where $t_1$ and $t_2$ are the times at which the particle enters and leaves
the cell, respectively.
While we do not have an analytic expression for the trajectory of the particle
(except when the particle follows a circular orbit), we can
numerically integrate the first-order ordinary
differential equations that
govern the particle's motion to a precision that is much higher than that of the partial
differential equation governing $\psi$. In this sense we treat the
integral over the source term as exact.
To evaluate the integral we adopt a four-point
Gauss-Legendre scheme, which has an error of order $h^8$.

\subsection{Potential term}
The most problematic term---from the point of view of implementing an
approximation of sufficiently high order in $h$---turns out to be the term
$V\psi$ in Eq.~\eqref{eqn:reduced-wave-eqn}. Since
this term does not contain a $\Dirac$-function, we have to approximate the
double integral 
\begin{align}
    {\int\!\!\int}_{\text{cell}} V \psi \,\d u\,\d v\label{eqn:pot-integral}
\end{align}
up to terms of order $h^6$ for a generic cell in order to achieve an overall $\orderof{h^4}$
convergence of the scheme. 

Here we have to treat cells traversed by the particle (``sourced'' cells)
differently
from the generic (``vacuum'') cells. While much of the algorithm can be transferred from
the vacuum cells to the sourced cells, some modifications are required. We
will describe each case separately in the following subsections.

\subsubsection{Vacuum case}
\begin{figure}[tbp]
    \centering
    \includegraphics{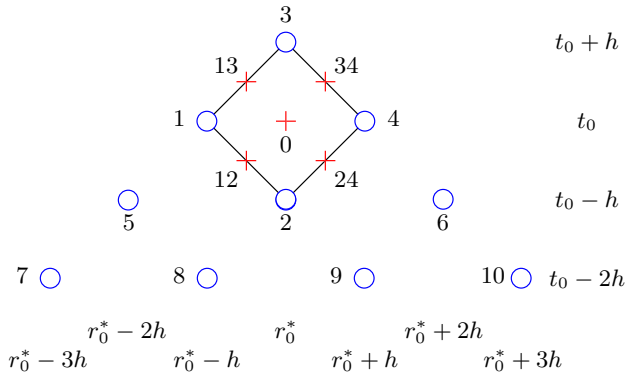}
    \caption{Points used to calculate the integral over the potential term for
    vacuum cells. Grid points are indicated by blue circles while red
    cross-hairs indicate points in between two grid points. 
    We calculate field values at points that do not lie on
    the grid by employing the second-order algorithm described
    in~\cite{lousto:05}.}
    \label{fig:simpson-points}
\end{figure}

To implement Lousto's algorithm to evolve the field across the vacuum cells, 
we use 
a double Simpson rule to compute the integral Eq.~\eqref{eqn:pot-integral}. 
We introduce the notation 
\begin{align}
    g(t,r^*) &= V(r^*) \, \psi(t, r^*)
\end{align}
and label our points in the same manner (see Fig.~\ref{fig:simpson-points}) as
in~\cite{lousto:05}:
\begin{align}
    {\int\!\!\int}_{\text{cell}} g \,\d u\,\d v =& \left(\frac{h}{3}\right)^2 [g_1 + g_2
    + g_3 + g_4 + 4(g_{12} + 
    \nonumber \\ & \mbox{} 
    g_{24} + g_{34} + g_{13}) + 16 g_0] + \orderof{h^6}.
    \label{eqn:simpson-rule}
\end{align}
Here, for example, $g_1$ is the value of $g$ at the grid point labeled $1$,
and $g_{12}$ is the value of $g$ at the off-grid point labeled $12$, etc. 
Deviating from Lousto's algorithm, we choose to calculate $g_0$ using
an expression different from that derived in~\cite{lousto:05}.
Unlike Lousto's approach, our expression exclusively involves points that are 
within the past light cone of the current cell. We find
\begin{align}
   g_0 =&
    \frac{1}{16} \bigl[\bigr.
      8 V_4 \, \psi_4
    + 8 V_1 \, \psi_1
    + 8 V_2 \, \psi_2 
    - 4 V_6 \, \psi_6
    - 4 V_5 \, \psi_5 
      \nonumber\\\mbox{}&
    + V_{10} \, \psi_{10}
    + V_7 \, \psi_7
    - V_9 \, \psi_9 
    - V_8 \, \psi_8
    \bigl. \bigr] + \orderof{h^4} \text{.}\label{eqn:g-naught}
\end{align}

In order to evaluate the term in parentheses in Eq.~\eqref{eqn:simpson-rule},
we again use a variant of the equations given
in~\cite{lousto:05}. Lousto's equations~(33) and~(34), 
\begin{align}
    g_{13} + g_{12} =& V(r_0^* - h/2) \, (\psi_1 + \psi_0) 
    \nonumber \\
    &\mbox{}\times
    \left[1 - \frac12 \left(\frac{h}{2}\right)^2 V(r_0^*-h/2)\right] + \orderof{h^4} 
    \text{,} \\
    g_{24} + g_{34} =& V(r_0^* + h/2) \, (\psi_0 + \psi_4) 
    \nonumber \\
    &\mbox{}\times
    \left[1 - \frac12 \left(\frac{h}{2}\right)^2 V(r_0^*+h/2)\right] + \orderof{h^4} 
\end{align}
contain isolated occurrences
of $\psi_0$, the value of the field at the central point. Since
Eq.~\eqref{eqn:g-naught} only allows us to find 
$g_0 = V_0 \psi_0$,
finding $\psi_0$ would involve a division by $V_0$, which will be numerically
unstable very close to the event horizon where $V_0 \approx 0$. Instead
we choose to express the potential term appearing in the square brackets as
a Taylor series around $r_0^*$. This allows us to eliminate the isolated
occurrences of $\psi_0$, and we find
\begin{align}
    g_{13} + &g_{12} + g_{24} + g_{34} =
     2V(r_0^*)   \,\psi_0\left[1 - \frac12 \left(\frac{h}{2}\right)^2 V(r_0^*)\right]
     \nonumber\\\mbox{}&
     +V(r_0^*-h/2)\,\psi_1\left[1 - \frac12 \left(\frac{h}{2}\right)^2 V(r_0^*-h/2)\right]
     \nonumber\\\mbox{}&
     +V(r_0^*+h/2)\,\psi_4\left[1 - \frac12 \left(\frac{h}{2}\right)^2 V(r_0^*+h/2)\right]
     \nonumber\\\mbox{}&
     +\frac12\bigl[V(r_0^*-h/2)-2V(r_0^*)
     \nonumber\\\mbox{}&
     +V(r_0^*+h/2)\bigr]\,(\psi_1+\psi_4) 
     + \orderof{h^4}
    \text{.}\label{eqn:sum-of-g}
\end{align}
Because of the $\left(\frac{h}{3}\right)^2$ factor in
Eq.~\eqref{eqn:simpson-rule}, this allows us to reach the required
$\orderof{h^6}$ convergence for a generic vacuum cell. This---given that
there is a number of order $N = 1/h^2$ of such cells---yields the desired overall
$\orderof{h^4}$ convergence of the full algorithm, at the end of the $N$ steps
required to finish the simulation.

\subsubsection{Sourced cells}\label{sec:sourced-cells}
For vacuum cells, the algorithm described above is the complete algorithm used to evolve the
field forward in time. For cells traversed by the particle, however, we have to
reconsider the assumptions used in deriving Eqs.~\eqref{eqn:g-naught}
and~\eqref{eqn:sum-of-g}. When deriving Eq.~\eqref{eqn:sum-of-g} we have employed the 
second-order evolution algorithm (see~\cite{lousto:05}), in which the single
step equation
\begin{align}
    \psi_3 =& -\psi_2 + \left(1-\frac{h^2}{2} V_0\right) 
	\left(\psi_1 + \psi_4\right)
\end{align}
is accurate only to
$\orderof{h^3}$ for cells traversed by the particle.
For these cells, therefore, the error term in
Eq.~\eqref{eqn:sum-of-g} is $\orderof{h^3}$ instead of $\orderof{h^4}$. As
there is a number of order $N' = 1/h$ of cells that are traversed by 
the particle in a
simulation run, the overall error---after including the
$\left(\frac{h}{3}\right)^2$ factor in Eq.~\eqref{eqn:simpson-rule}---is of
order $h^4$.
We can therefore afford this reduction of the convergence order in
Eq.~\eqref{eqn:sum-of-g}

Equation~\eqref{eqn:g-naught}, however, is accurate only to $\orderof{h}$ for 
cells traversed by the particle. Again taking
the $\left(\frac{h}{3}\right)^2$ factor into account, this renders the 
overall algorithm $\orderof{h^2}$.
\begin{figure}[tbp]
    \centering
    \includegraphics{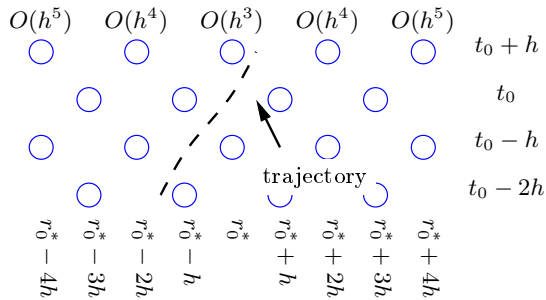}
    \caption{Cells affected by the passage of the particle, showing the
    reduced order of the single step equation}\label{fig:affected-cells}
\end{figure}
Figure~\ref{fig:affected-cells} shows the cells affected by the particle's
traversal and the reduced order of the single step equation for each cell.
Cells whose convergence order is $\orderof{h^5}$ or higher do not need modifications,
since there is only a number $N'=1/h$ of such cells in the simulation. We
are therefore concerned about cells neighboring the particle's
trajectory and those traversed by the particle. 

\paragraph{Cells neighboring the particle}
These cells are not traversed by the particle, but the particle
might have traversed cells in their past light-cone, which are used in the
calculation of $g_0$ in Eq.~\eqref{eqn:g-naught}. For these cells, we use a
one-dimensional Taylor expansion of $g(t, r^*)$ within the current time-slice
$t = t_0$,
\begin{align}
    g_0 =& \frac{1}{16} \bigl[ 
       5 V(r_0^* -  h)\,\psi(t_0, r_0^* -  h)
     \nonumber\\\mbox{}&
    + 15 V(r_0^* - 3h)\,\psi(t_0, r_0^* - 3h)
     \nonumber\\\mbox{}&
    -  5 V(r_0^* - 5h)\,\psi(t_0, r_0^* - 5h)
     \nonumber\\\mbox{}&
    +    V(r_0^* - 7h)\,\psi(t_0, r_0^* - 7h)
    \bigr]
    + \orderof{h^4}
     \label{eqn:g-naught-left}
\end{align}
for the cell on the left-hand side, and
\begin{align}
    g_0 =& \frac{1}{16} \bigl[ 
       5 V(r_0^* +  h)\,\psi(t_0, r_0^* +  h)
     \nonumber\\\mbox{}&
    + 15 V(r_0^* + 3h)\,\psi(t_0, r_0^* + 3h)
     \nonumber\\\mbox{}&
    -  5 V(r_0^* + 5h)\,\psi(t_0, r_0^* + 5h)
     \nonumber\\\mbox{}&
    +    V(r_0^* + 7h)\,\psi(t_0, r_0^* + 7h)
    \bigr]
    + \orderof{h^4}
     \label{eqn:g-naught-right}
\end{align}
for the cell on the right-hand side,
where $(t_0, r_0^*)$ is the center of the cell traversed by the particle.
Both of these are more accurate than is strictly necessary; we would need
error terms of order $h^3$ to achieve the desired overall
$\orderof{h^4}$ convergence of the algorithm. Keeping the extra terms,
however,  improves the numerical convergence slightly.

\paragraph{Cell traversed by the particle}
We
choose
not to implement a fully explicit algorithm to handle cells traversed by the
particle, because this would
increase the complexity of the algorithm by a significant factor. 
Instead we use an
iterative approach to evolve the field using the integrated wave equation 
\begin{align}
   -4(\psi_3 + &\psi_2 - \psi_1 - \psi_4)
   - {\int\!\!\int}_{\text{cell}} V\,\psi \,\d u\,\d v = 
   \nonumber\\\mbox{}&
   -\frac{8\pi q}{E} 
   \int_{t_1}^{t_2} \frac{f_0(t)}{r_0(t)} \conj{Y}_{\ell m}\bm(\pi/2, \phi_0(t)\bm) \,\d t \text{.}
\end{align}
In this equation the integral involving the source term can be evaluated to
any desired accuracy at the beginning of the iteration, because the motion of
the particle is determined by a simple system of ordinary differential
equations, which are easily integrated with reliable numerical methods.
It remains to evaluate the integral over the potential term, which we do
iteratively.  Schematically the method works as follows:
\begin{itemize}
    \item Make an initial guess for $\psi_3$ using the second-order
	algorithm. This guess is correct up to terms of $\orderof{h^3}$.
    \item Match a second-order piecewise interpolation polynomial 
	to the six points
        that make up the past light-cone of the future grid point, including
        the future point itself.
        \begin{figure}[tbp]
            \centering
    \includegraphics{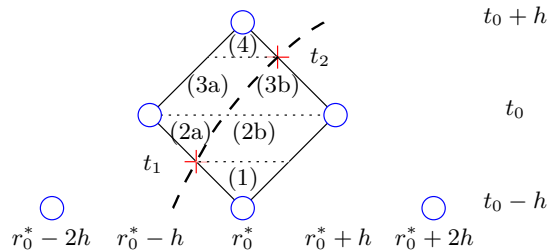}
            \caption{Typical cell traversal of the particle. We split
            the domain into sub-parts indicated by the dotted line
            based on the time the particle enters (at $t_1$) and
            leaves (at $t_2$) the cell. The integral over each
            sub-part is evaluated using an iterated two-by-two point
            Gauss-Legendre rule.}\label{fig:potint-subparts}
        \end{figure}
    \item Use this approximation for $\psi$ to numerically calculate
        \begin{align*}
	    {\int\!\!\int}_{\text{cell}} V\,\psi \,\d u\,\d v \text{,}
        \end{align*}
        using two-by-two point Gauss-Legendre rules for the six sub-parts
        indicated in Fig.~\ref{fig:potint-subparts}. 
    \item Update the future value of the field and repeat the process
        until the iteration has converged to a required degree of accuracy.
\end{itemize}

\section{Initial values and boundary conditions}\label{sec:initial-and-boundary-condition}
As is typical for numerical simulations, we have to pay careful attention to
specifying initial data and appropriate boundary conditions. 
These aspects of the numerical method are highly non-trivial problems
in full numerical relativity, but they can be solved or circumvented with moderate
effort in the present work.

\subsection{Initial data}
\begin{figure}[tbp]
    \centering
    \includegraphics{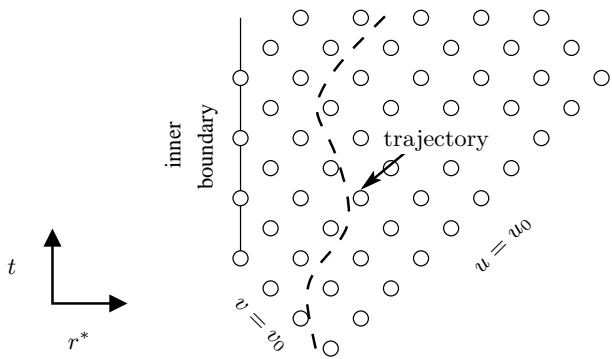}
    \caption{Numerical domain evolved during the simulation. 
    We impose an inner boundary condition close to the black whole where
    we can implement it easily to the accuracy of the underlying floating
    point format. Far away from the black hole, we evolve the full domain of
    dependence of the initial data domain without imposing boundary conditions.}
    \label{fig:numerical-domain}
\end{figure}
In this work we use a characteristic grid consisting of points lying on
characteristic lines of the wave operator to evolve $\psi$ forward in time.
As such, we need to specify characteristic initial data on the lines $u = u_0$
and $v = v_0$ shown in Fig.~\ref{fig:numerical-domain}.
We choose not to worry about specifying ``correct'' initial
data, but instead arbitrarily choose $\psi$ to vanish on $u =
u_0$ and $v = v_0$:
\begin{align}
    \psi(u = u_0, v) = \psi(u, v=v_0) = 0 \text{.}
\end{align}
This is equivalent to adding spurious initial waves in the form of a
homogeneous solution of Eq.~\eqref{eqn:reduced-wave-eqn} to the correct
solution. This produces an initial wave burst that moves away from the particle
with the speed of light, and quickly leaves the numerical domain. Any
remaining tails of the spurious initial data decay as $t^{-(2\ell + 2)}$ as
shown in~\cite{price:72} and become negligible after a short time.
We conclude that the influence of the initial-wave content on the self-force
becomes negligible after a time of the order of the light-crossing
time of the particle's orbit. 

\subsection{Boundary conditions}
On the analytical side we would like to impose ingoing boundary conditions at
the event horizon $r^* \approaches -\infty$ and outgoing boundary conditions
at spatial infinity $r^* \approaches \infty$, ie.
\begin{align}
    \lim_{r^* \approaches -\infty} \partial_u \psi =& 0\text{,} & 
    \lim_{r^* \approaches  \infty} \partial_v \psi =& 0
    \text{.}
\end{align}
Because of the finite resources available to a computer we can only simulate a
finite region of the spacetime, and are faced with the reality of
implementing boundary conditions at finite values of $r^*$. 
Two solutions to this problem present themselves:
\begin{enumerate}
    \item choose the numerical domain to be the domain of dependence of the
	initial data surface. Since
	the effect of the boundary condition can only propagate forward in
	time with at most the speed of light, this
	effectively hides any influence of the boundary. This is what we choose
	to do in order to deal with the outer boundary condition.
    \item implement boundary conditions sufficiently ``far out'' so that
	\emph{numerically} there is no difference between imposing the
	boundary condition there or at infinity. Since the boundary conditions
	depend on the vanishing of the potential $V(r)$
	appearing in the wave equation, this will happen once $1-2M/r \approx
	0$. Near the horizon 
	$r \approx 2 M (1 + \exp(r^*/2 M))$,
	so this will happen---to numerical accuracy---for modestly large 
	(negative) values of 
	$r^* \approx -73\,M$. We choose to implement the ingoing waves
	condition $\partial_u \psi_{\ell m} = 0$ there.
\end{enumerate} 
\section{Implementation}\label{sec:implementation}
Making more precise the ideas developed in the preceding sections, we implement
the following numerical scheme. 

\subsection{Particle motion}\label{sec:particle-motion}
Following Darwin~\cite{darwin:59} we introduce the dimensionless semi-latus
rectum $p$ and the eccentricity $e$ such that for a bound orbit around a
Schwarzschild black hole of mass $M$,
\begin{align}
    r_1 &= \frac{pM}{1+e}\text{,} & r_2 &= \frac{pM}{1-e}
\end{align}
are the radial positions of the periastron and apastron, respectively. 
Energy per unit mass
and angular momentum per unit mass are then given by
\begin{align}
    E^2 &= \frac{(p-2-2e)(p-2+2e)}{p\,(p-3-e^2)}\text{,} & 
    L^2 &= \frac{p^2M^2}{p-3-e^2} \text{.}
\end{align}
Together with these definitions it is useful to introduce an orbital parameter
$\chi$ such that along the trajectory of the particle,
\begin{align}
    r(\chi) &= \frac{pM}{1+e\cos\chi}\text{,}\label{eqn:chi-def}
\end{align}
where $\chi$ is single-valued along the orbit. We can then write down 
first-order differential equations for $\chi(t)$ and the azimuthal angle
$\phi(t)$ of the particle,
\begin{align}
    \odiff{\chi}{t} =&
    \frac{(p-2-2e\cos\chi)(1+e\cos\chi)(1+e\cos\chi)}{(Mp^2)}
    \nonumber\\
    \mbox{}&\times
    \sqrt{ \frac{p-6-2e\cos\chi}{(p-2-2e)(p-2+2e)} }\text{,} \\
    \odiff{\phi}{t} =&
    \frac{(p-2-2e\cos\chi)(1+e\cos\chi)^2}{p^{3/2}M\sqrt{(p-2-2e)(p-2+2e)}}
    \text{.}\label{eqn:particle-ode}
\end{align}
We use the embedded Runge-Kutta-Fehlberg (4, 5) algorithm provided by the
\programname{GNU Scientific Library} routine
\routinename{gsl\_odeiv\_step\_rkf45} and an adaptive step-size control to
evolve the position of the particle forward in time. Intermediate values of
the particle's position are found using a Hermite interpolation of the nearest
available calculated positions.

\subsection{Initial data}
We do not specify initial data. The field is set to zero on the 
initial characteristic slices, $u =u_0$ and $v = v_0$. 

\subsection{Boundary conditions}
We adjust the outer boundary of the numerical domain at each time-step so 
that we cover the domain of dependence of the initial characteristic surfaces
and the particle's world line.
The resulting numerical domain was already shown in Fig.~\ref{fig:numerical-domain}.

Near the event horizon, at $r^* \approx -73\,M$, we implement an ingoing-wave
boundary condition by imposing
\begin{align}
    \psi(t+h, r^*) &= \psi(t,r^*-h) \text{.}
\end{align}
This allows us to drastically reduce the number of cells in the
numerical domain, and consequently the running time of the simulation.

\subsection{Evolution in vacuum}
Cells not traversed by the particle are evolved using
Eqs.~\eqref{eqn:differential-operator}, \eqref{eqn:simpson-rule} -- \eqref{eqn:sum-of-g}.
Explicitly written out, we use
\begin{align}
    &\psi_3 = -\psi_2 
      \nonumber\\\mbox{}&
      + \biggl[
        1 -
        \frac14\left(\frac{h}{3}\right)^2\,(V_0+V_1) 
        + \frac{1}{16}\left(\frac{h}{3}\right)^4 V_0\,(V_0+V_1)
	\biggr]\psi_1 
      \nonumber\\\mbox{}&
      + \biggl[
        1 -
        \frac14\left(\frac{h}{3}\right)^2\,(V_0+V_4)
        + \frac{1}{16}\left(\frac{h}{3}\right)^4 V_0\,(V_0+V_4)
	\biggr]\psi_4
    \nonumber\\\mbox{}&
      - \biggl[1 -
      \frac14\left(\frac{h}{3}\right)^2V_0\biggr]\left(\frac{h}{3}\right)^2
      (g_{12}+g_{24}+g_{34}+g_{13}+4g_0)
      \text{,}\label{eqn:vacuum-evolution}
\end{align}
where $g_0$ is  
given by Eq.~\eqref{eqn:g-naught} and the sum $g_{12}+g_{24}+g_{34}+g_{13}$
is given by Eq.~\eqref{eqn:sum-of-g}.

\subsection{Cells next to the particle}
Vacuum cells close to the current position of the particle require a different approach 
to calculate $g_0$, since the cells in their past light cone could have been traversed by the particle. 
We use Eqs.~\eqref{eqn:g-naught-left} and~\eqref{eqn:g-naught-right} to find
$g_0$ in this case.
Other than this modification, the same algorithm as for generic vacuum cells is used.

\subsection{Cells traversed by the particle}
We evolve cells traversed by the particle using the iterative algorithm described
in Sec.~\ref{sec:sourced-cells}. Here
\begin{align}
    \psi_3 =&  - \psi_1 + \psi_2 
    \nonumber\\ \mbox{}
        & + \psi_4
	-\frac14 {\int\!\!\int}_{\text{cell}} V\,\psi \,\d u\,\d v
	\nonumber\\ \mbox{}&
	+ \frac{2\pi q}{E} \int_{t_1}^{t_2} \frac{f_0(t)}{r_0(t)}
	\conj{Y}_{\ell	m}\bm(\pi/2, \pi_0(t)\bm) \, \d t \text{,}
\end{align}
where the initial guess for the iterative evolution of
$\int\!\!\int_{\text{cell}} V \psi \,\d u\,\d v$ is obtained using the second
order algorithm of Lousto and Price~\cite{lousto:97},
\begin{align}
    \psi_3 =& -\psi_1 + \left[1-\frac{h^2}{2} V_0\right] 
	\nonumber\\ \mbox{}&\times
	\left[\psi_2 + \psi_4\right] 
	\nonumber\\\mbox{}&
	+ \frac{2\pi q}{E} \int_{t_1}^{t_2} \frac{f_0(t)}{r_0(t)}
	\conj{Y}_{\ell	m}\bm(\pi/2, \pi_0(t)\bm) \, \d t 
    	\text{.}\label{eqn:second-order-sourced}
\end{align}
Successive iterations use a four-point Gauss-Legendre rule to evaluate the
integral of $V\psi$; this requires a second-order polynomial interpolation of the
current field values as described in Appendix~\ref{sec:piecewise-polynomials}.

\subsection{Extraction of the field data at the
particle}\label{sec:field-extraction}
In order to extract the value of the field and its first derivatives at the
position of the particle, we again use a polynomial interpolation
at the points surrounding the particle's position. Using a fourth-order
polynomial, as described in Appendix~\ref{sec:piecewise-polynomials}, we can
estimate $\psi$, $\partial_t \psi_t$, and $\partial_{r^*} \psi$ at the position
of the particle up to errors of order $h^4$. As was briefly mentioned in
Sec.~\ref{sec:numerical-method}, we would expect an error term of order
$h^3$ for $\partial_t \psi_t$ and $\partial_{r^*} \psi$. 
The $\orderof{h^4}$ accuracy we actually achieve by using a fourth-order 
(instead of a third-order)
piecewise polynomial shows up clearly in a regression plot such as
Fig.~\ref{fig:sourced-regression-derivative}.

\subsection{Regularization of the mode sum}
We use the calculated multipole moments $\psi_{\ell m}$ to construct the
multipole moments $\Phi_{\ell m}$, and first derivatives $\partial_t \Phi_{\ell m}$ and 
$\partial_r \Phi_{\ell m}$, of the scalar field. These, in turn, are used to calculate the tetrad
components $\Phi_{(0)\ell m}$, $\Phi_{(+)\ell m}$, $\Phi_{(-)\ell m}$, and $\Phi_{(3)\ell m}$ of the
field gradient according to Eqs.~(1.23)--(1.26) of~\cite{haas:06}, which are
reproduced in Appendix~\ref{sec:translation-tables}.
These multipoles then give rise to the multipole coefficients of the
retarded field,
\begin{align}
    \Phi_{(\mu)\ell}(t,r,\theta,\phi) =& 
    \sum_{m=-\ell}^\ell \Phi_{(\mu)\ell m}(t,r)Y_{\ell m}(\theta,\phi)\text{,}
\end{align}
which are subjected to the regularization procedure described by 
Eq.~(1.29) of~\cite{haas:06},
\begin{align}
    \Phi^{\mathrm{R}}_{(\mu)}(t, r_0, \pi/2, \phi_0)  
    =& \lim_{\Delta \approaches 0} \sum_\ell \biggl\{ 
    \Phi_{(\mu)\ell}(t, r_0+\Delta, \pi/2, \phi_0) 
    \nonumber\\ \mbox{}&
    - q \bigl[ (\ell + 1/2) A_{(\mu)} 
    + B_{(\mu)} 
    \nonumber\\ \mbox{}&
    + \frac{C_{(\mu)}}{(\ell + 1/2)}
    + \frac{D_{(\mu)}}{(\ell - 1/2) 
    (\ell + 3/2)} 
    \nonumber\\ \mbox{}&
    + \cdots \bigr]
    \biggr\}\text{,}\label{eqn:regularize-field}
\end{align}
using the regularization parameters $A_{(\mu)}$, $B_{(\mu)}$, $C_{(\mu)}$, and
$D_{(\mu)}$ tabulated in Appendix~\ref{sec:regularization-parameters}.

Finally we reconstruct the vector components of the field gradient 
using Eqs.~(1.47)--(1.48) of~\cite{haas:06},
\begin{align}
    \Phi^{\mathrm{R}}_t      &= \sqrt{f_0} \Phi^{\mathrm{R}}_{(0)}, \\
    \Phi^{\mathrm{R}}_r      &= \frac{1}{\sqrt{f_0}} 
    \left( \Phi^{\mathrm{R}}_{(+)} e^{-i\phi_0} 
	+ \Phi^{\mathrm{R}}_{(-)} e^{i\phi_0} \right), \\
    \Phi^{\mathrm{R}}_\theta &= -r_0 \Phi^{\mathrm{R}}_{(3)}, \\
    \Phi^{\mathrm{R}}_\phi   &= -\frac{i r_0}{2} 
    \left( \Phi^{\mathrm{R}}_{(+)} e^{-i\phi_0} 
	- \Phi^{\mathrm{R}}_{(-)} e^{i\phi_0} \right),
\end{align}
and calculate the self-force
\begin{align}
    F_\alpha &= q \Phi^R_\alpha\text{.}
\end{align}

We recall the discussion in Sec.~\ref{sec:the-problem}
concerning the definition of $\Phi^R$, its connection to the self-force acting
on the particle, and its regularity at the particle's position. 
    
\section{Numerical tests}\label{sec:numerical-tests}
In this section we present the tests we have performed to validate
our numerical evolution code. First, in order to check the fourth-order
convergence rate of the code, we perform regression runs with increasing
resolution for both a vacuum test case, where we seeded the evolution
with a Gaussian wave packet, and a case where a particle is
present. As a second test, we compute the regularized self-force for several
different combinations of orbital elements $p$ and $e$ and check that the
multipole coefficients decay with $\ell$ as expected. This provides a very
sensitive check on the overall implementation of the numerical scheme, as well
as the analytical calculations that lead to the regularization parameters.
Finally, we calculate the self-force for a particle on a circular orbit and
show that it agrees with the results presented 
in~\cite{haas:06, diaz-rivera:04}.

\subsection{Convergence tests: Vacuum}\label{sec:convergence-tests-vacuum}
\begin{figure}[tbp]
    \centering
    \includegraphics{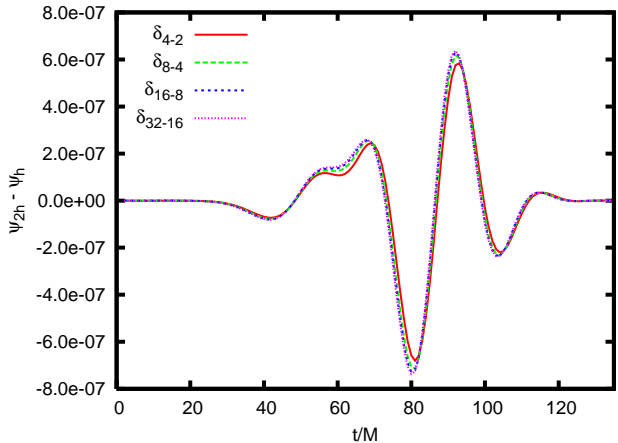}
    \caption{Convergence test of the numerical algorithm in the vacuum case. We
    show differences between simulations using different step sizes 
    $h = 0.5\,M$
    ($\psi_2$), $h = 0.25\,M$ ($\psi_4$), $h = 0.125\,M$ ($\psi_8$), $h =
    0.0625\,M$ ($\psi_{16}$), and $h = 0.03125\,M$ ($\psi_{32}$).
    Displayed are the rescaled differences $\delta_{4-2} = \psi_4 - \psi_2$,
    $\delta_{8-4} = 2^4(\psi_8 - \psi_4)$, $\delta_{16-8} = 4^4(\psi_8 - \psi_4)$, 
    and $\delta_{32-16} = 8^4(\psi_8 - \psi_4)$ for the real part of the 
    $\ell = 2$, $m = 2$ mode at $r^* \approx 20\,M$. The maximum value of the field
    itself is of the order of $0.1$, so that the errors in the field values
    are roughly five orders
    of magnitude smaller than the field values themselves.
    We
    can see that the convergence is in fact of fourth-order, as the curves
    lie nearly on top of each other, with only the lowest resolution curve
    $\delta_{4-2}$ deviating slightly.}\label{fig:vacuum-regression}
\end{figure}

As a first test of the validity of our numerical code we estimate the
convergence order by removing the particle and performing regression runs for
several resolutions.
We use a Gaussian wave packet as initial data,
\begin{align}
    \psi(u = u_0, v) &= \exp(-[v-v_p]^2/[2\sigma^2])\text{,} \\
    \psi(u, v = v_0) &= 0 \text{,}
\end{align}
where $v_p = 75\,M$ and $\sigma = 10\,M$, $v_0 = -u_0 = 6\,M + 2\,M \ln2$,
and we extract the field values at
$r^* = 20\,M$. Several such runs were performed, 
with varying resolution of 2, 4, 8, 16,
and 32 grid points per $M$. Figure~\ref{fig:vacuum-regression} shows
$\psi(2h) - \psi(h)$ rescaled by appropriate powers of $2$, so
that in the case of fourth-order convergence the curves would lie on top of
each other. As can be seen from the plots, they do, and the vacuum portion of
the code is indeed fourth-order convergent. 
\subsection{Convergence tests: Particle}\label{sec:convergence-tests-particle}
\begin{figure}[tbp]
    \centering
    \includegraphics{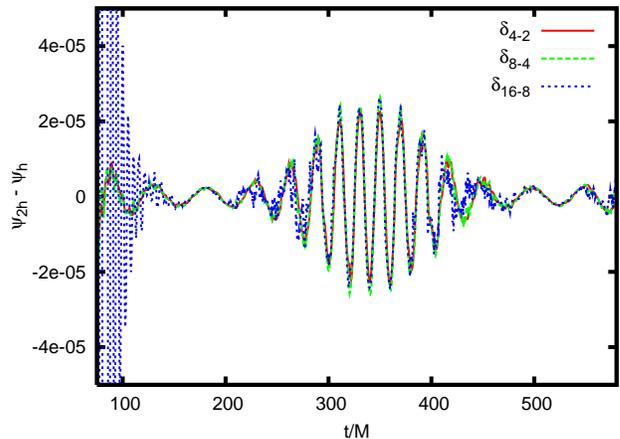}
    \caption{Convergence test of the numerical algorithm in the sourced case. We
    show differences between simulations using different step sizes of 
    4 ($\psi_4$), 8 ($\psi_8$), 16 ($\psi_{16}$), and 32
    ($\psi_{32}$) cells per $M$.
    Displayed are the rescaled differences $\delta_{8-4} = \psi_8 - \psi_4$,  etc.\ 
    (see caption of Fig.~\ref{fig:vacuum-regression} for definitions)
    of the field values at the position of
    the particle for a simulation with $\ell = 6$, $m = 4$ and $p = 7$, $e = 0.3$.  
    We see that the convergence is approximately fourth-order.}
    \label{fig:sourced-regression}
\end{figure}

While the convergence test described in
section~\ref{sec:convergence-tests-vacuum} clearly shows that the desired
convergence is achieved for vacuum evolution, it does not
test the parts of the code that are used in the integration of the 
inhomogeneous
wave equation. To test these  
we perform a second set of
regression runs, this time using a non-zero charge $q$. We extract 
the field at the position of the particle, thus also testing the
implementation of the extraction algorithm described in
section~\ref{sec:field-extraction}. 
For this test we choose the $\ell = 6$, $m = 4$ mode of the field generated by
a particle on a mildly eccentric
geodesic orbit with $p = 7$, $e = 0.3$. As shown in
Fig.~\ref{fig:sourced-regression} the convergence is still of fourth
order, but the two curves no longer lie precisely on top of each other
at all times.
\begin{figure}[tbp]
    \centering
    \includegraphics{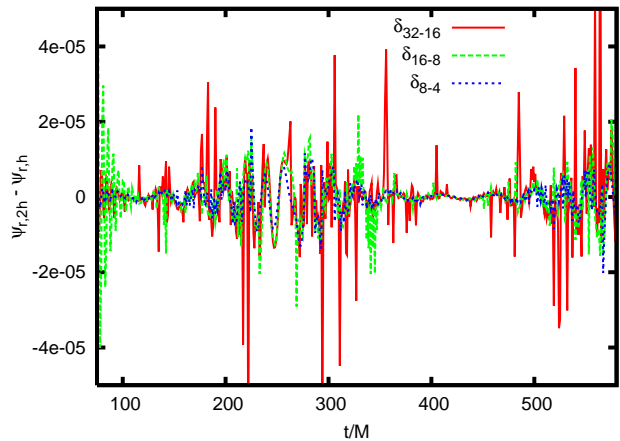}
    \caption{Convergence test of the numerical algorithm in the sourced case. We
    show differences between $\partial_r \Phi$ for simulations using different 
    step sizes of 
    4 ($\Phi_{r,4}$), 8 ($\Phi_{r,8}$), 16 ($\Phi_{r,16}$), and 32
    ($\Phi_{r,32}$) cells per $M$.
    Displayed are the rescaled differences $\delta_{8-4} = \Phi_{r,8} -
    \Phi_{r,4}$ etc.\ 
    of the values at the position of
    the particle for a simulation with $\ell = 6$, $m = 4$ and $p = 7$, $e = 0.3$.  
    Although there is much noise caused by the piecewise polynomials used to
    extract the data, we can see that the convergence is 
    approximately fourth-order.}
    \label{fig:sourced-regression-derivative}
\end{figure}
The region before $t \approx 100\,M$ is dominated by the initial wave burst
and therefore does not scale as expected, yielding two very different curves.
In the region $300\,M \lesssim t \lesssim 400\,M$ the two curves lie on top of
each other, as expected for a fourth-order convergent algorithm. In the region
between $t \approx 200\,M$ and $t \approx 300\,M$, however, the dashed curves
have slightly smaller amplitudes than the solid one, indicating an
order of convergence different from (but close to) four. 

To explain this behavior we have to 
examine
the terms that contribute significantly to the error in the simulation. 
The numerical error is almost completely dominated by that of the 
approximation of the potential term
$\int\!\!\int_{\text{cell}} V\psi \, \d u \, \d v$ in the integrated wave 
equation. For vacuum cells the error in this approximation 
scales as $h^6$, where $h$ is the
step size. For cells traversed by the particle, on the other hand, the
approximation error depends also on the difference $t_2 - t_1$ of the times 
at which the particle enters and leaves the cell. This difference is bounded by $h$
but does not necessarily scale as $h$. For example, if a particle enters a cell 
at its
very left, then scaling $h$ by $\frac12$ would not change $t_2 - t_1$
at all, thus leading to a scaling behavior that differs from expectation.

\begin{figure}[tbp]
    \centering
    \includegraphics{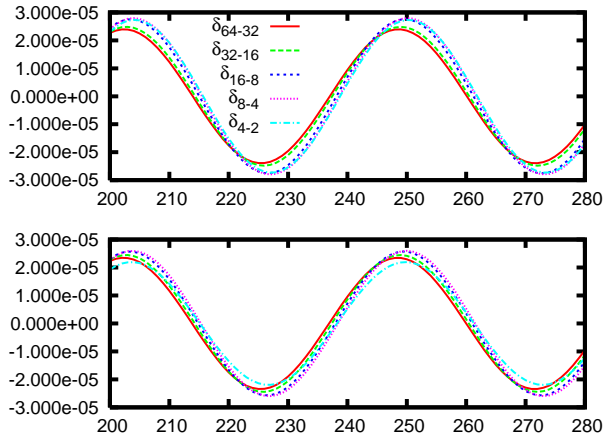}
    \caption{Behavior of convergence tests for a particle in circular orbit at
    $r = 6\,M$. We show differences between simulations of the $\ell = 2$, $m
    = 2$ multipole moment using different 
    step sizes of 2 ($\psi_2$), 4 ($\psi_4$), 
    8 ($\psi_8$), 16 ($\psi_{16}$), 32 ($\psi_{32}$) and 
    64 ($\psi_{64}$) cells per $M$.
    Displayed are the real part of the rescaled
    differences $\delta_{4-2} = (\psi_{4} - \psi_{2})$ etc.\ 
    of the field values at the position of
    the particle, defined as in Fig.~\ref{fig:vacuum-regression}.
    The values have been rescaled so that---for fourth order
    convergence---the curves should all coincide. 
    The upper panel corresponds to a set of simulations
    where the particle traverses the cells away from their tips.
    The curves do not coincide perfectly with each other, seemingly indicating 
    a failure of the
    convergence. The lower panel was obtained in a simulation where
    the particle was carefully positioned so as to pass through the tips of each
    cell it traverses. This set of simulations passes the
    convergence test more convincingly.}
    \label{fig:convergence-order-position}
\end{figure}

\begin{figure}[tbp]
    \centering
    \includegraphics{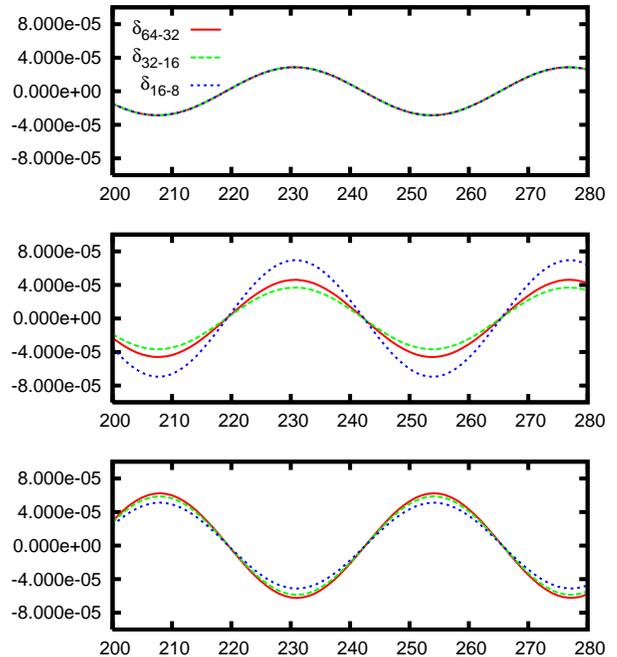}
    \caption{Behavior of convergence tests for a particle in circular orbit at
    $r = 6\,M$. We show differences between simulations of the $\ell = 2$, $m
    = 2$ multipole moment using different 
    step sizes of 8 ($\psi_8$), 16 ($\psi_{16}$), 
    32 ($\psi_{32}$), and 
    64 ($\psi_{64}$) cells per $M$.
    Displayed are the real part of the rescaled
    differences $\delta_{16-8} = \psi_{16} - \psi_{8}$ etc.\ 
    of the field values at the position of
    the particle, defined as in Fig.~\ref{fig:vacuum-regression}.
    The values have been rescaled so that---for second order
    convergence---the curves should 
    all coincide. The upper two panels correspond to 
    simulations where the second order algorithm was used throughout.
    For the topmost one, care was taken to ensure that the particle passes
    through the tip of each cell it traverses, while in the middle one no such
    precaution was taken.
    Clearly the curves in the middle panel do not coincide with each other,
    indicating a failure of the second-order convergence of the code.
    The lower panel was obtained in a simulation using the
    mixed-order algorithm described in the text.
    While the curves still do not coincide precisely, the observed behavior is
    much closer to the expected one than for the purely second order
    algorithm.}
    \label{fig:convergence-order-accuracy}
\end{figure}

To investigate this further we conducted test runs of the simulation for a
particle on a circular orbit at $r = 6\,M$. In
order to observe the expected scaling behavior, we have to make sure that the
particle passes through the tips of the cell it traverses. When this is the
case, then $t_2 - t_1 \equiv h$ and a plot similar to the one shown in
Fig.~\ref{fig:sourced-regression} shows the proper scaling behavior. As a
further test we artificially reduced the convergence order of the
\emph{vacuum} algorithm to two by implementing the second-order
algorithm described in~\cite{lousto:05}. By keeping the algorithm that deals
with sourced cells unchanged, we reduced the relative impact on the numerical
error. This, too, allows us to recover the expected (second-order)
convergence. Figures~\ref{fig:convergence-order-position} and
\ref{fig:convergence-order-accuracy} illustrate the effects of the
measures taken to control the convergence behavior.

\subsection{High-$\ell$ behavior of the multipole coefficients}\label{sec:high-ell-behavior}
\begin{figure}[tbp]
    \centering
    \includegraphics{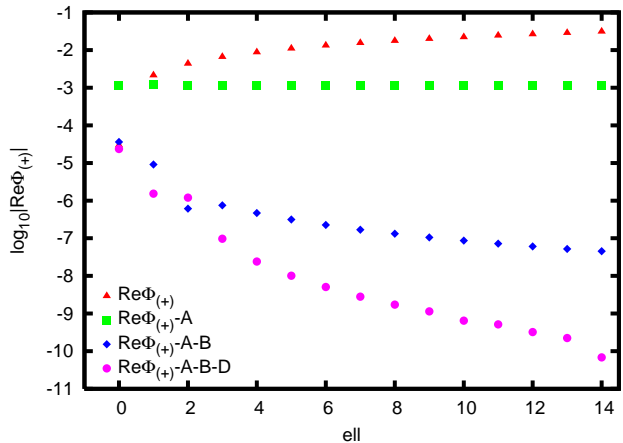}
    \caption{Multipole coefficients of the dimensionless self-force
    $\frac{M^2}{q} \real\Phi^{\mathrm{R}}_{(+)}$ for a
    particle on an eccentric orbit ($p = 7.2$, $e = 0.5$).
    The
    coefficients are extracted at $t = 500\,M$ along the trajectory shown in 
    Fig.~\ref{fig:eccentric-orbit-trajectory}.
    The plots show several stages of the
    regularization procedure, with a closer
    description of the curves to be found in the
    text.}\label{fig:multipole-coeffs-eccentric-orbit-plus}
\end{figure}

Inspection of Eq.~\eqref{eqn:regularize-field} reveals that a plot of
$\Phi_{(\mu)\ell}$ as a function of $\ell$ (for a selected value of $t$)
should display a linear growth in $\ell$ for large $\ell$. Removing the
$A_{(\mu)}$ term should produce a constant curve, removing the $B_{(\mu)}$
term (given that $C_{(\mu)} = 0$) should produce a curve that decays as
$\ell^{-2}$, and finally, removing the $D_{(\mu)}$ term should produce a curve
that decays as $\ell^{-4}$. It is a powerful test of the numerical methods to
check whether these expectations are borne out by the numerical data. 
Fig.~\ref{fig:multipole-coeffs-eccentric-orbit-plus}
plots the remainders as obtained from our numerical simulation,
demonstrating the expected behavior.
It displays, on a logarithmic scale, the absolute value of 
$\real\Phi^{\mathrm{R}}_{(+) \ell}$, the real part of the $(+)$ component of
the self-force.
The orbit is eccentric ($p =7.2$, $e = 0.5$), and all components of the self-force 
require regularization. The
first curve (in triangles) shows the unregularized multipole coefficients that
increase linearly in $\ell$, as confirmed by fitting 
a straight line to the data.
%
The second curve (in squares) shows partially regularized coefficients,
obtained after
the removal of $(\ell+1/2)A_{(\mu)}$; this  clearly approaches a constant for large
values of $\ell$. The curve made up of diamonds shows the behavior after
removal of $B_{(\mu)}$; because 
$C_{(\mu)} = 0$, it decays as $\ell^{-2}$, a behavior that is confirmed by a fit to
the $\ell \ge 5$ part of the curve.
%
Finally, after removal of $D_{(\mu)}/[(\ell - \frac12)\,(\ell + \frac32)]$ the
terms of the sum decrease in magnitude as $\ell^{-4}$ for large values of
$\ell$, as derived in~\cite{detweiler:03a}. 
%
Each one of the last two curves would result in a
converging sum, but the convergence is much faster after subtracting the
$D_{(\mu)}$ terms. We thereby gain more than 2 orders of magnitude in the
accuracy of the estimated sum.

Figure~\ref{fig:multipole-coeffs-eccentric-orbit-plus} provides a sensitive
test of
the implementation
of both the numerical and analytical parts of the calculation. Small
mistakes in either one will cause the difference in
Eq.~\eqref{eqn:regularize-field} to have a vastly different behavior.

\subsection{Self-force on a circular orbit}\label{sec:self-force-on-circular-orbit}
\begin{figure}[tbp]
    \centering
    \includegraphics{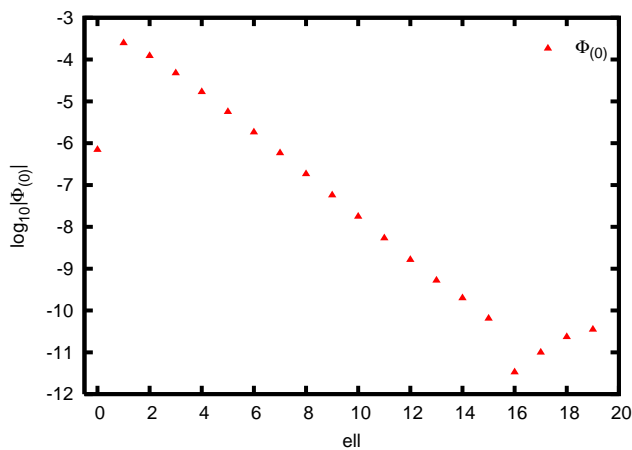}
    \caption{Multipole coefficients of $\Phi^{\mathrm{R}}_{(0)}$ for a particle
    on a circular orbit. Note that $\Phi^{\mathrm{R}}_{(0) \ell}$ is linked to
    $\Phi^{\mathrm{R}}_t$ via
    $\Phi^{\mathrm{R}}_t = \sqrt{f_0} \Phi^{\mathrm{R}}_{(0)}$. The multipole
    coefficients decay exponentially with $\ell$ until $\ell \approx 16$, at
    which point numerical errors start to dominate.}
    \label{fig:multipole-coeffs-circular-orbit-t}
\end{figure}

For the case of a circular orbit, the regularization parameters $A_{(0)}$,
$B_{(0)}$, and $D_{(0)}$ all vanish identically, so that the $(0)$ (or
alternatively the $t$) component of the self-force does not require
regularization. Figure~\ref{fig:multipole-coeffs-circular-orbit-t} thus 
shows only one curve, with the magnitude of the multipole coefficients
decaying exponentially with increasing $\ell$.

As a final test, in Table~\ref{tab:regul-force-circular-orbit} 
we compare our result for the self-force on a particle in a
circular orbit at $r = 6M$ to those obtained
in~\cite{haas:06, diaz-rivera:04} using a frequency-domain code. 
\begingroup
\squeezetable
\begin{table}
    \caption{Results for the self-force on a scalar particle with scalar charge
    $q$ on a circular orbit at $r_0 = 6M$. 
    The first column lists the results as calculated in this work using
    time-domain numerical methods, while
    the second and third columns list the results as calculated in~\cite{haas:06,
    diaz-rivera:04} using frequency-domain methods. 
    For the $t$ and $\phi$ components the number of digits is limited by 
    numerical roundoff error. For the $r$ component the number of digits is
    limited by the truncation error of the sum of
    multipole coefficients.}
    \begin{ruledtabular}
    \begin{tabular}{lccc}
	& This work:  & Previous work:                  & Diaz-Rivera \\
	& time-domain & frequency-domain \cite{haas:06} & et. al.\ \cite{diaz-rivera:04}\\
	\hline
	$\frac{M^2}{q}\Phi_{t}^R$    & $3.60339\times10^{-4}$  & $3.60907254\times10^{-4}$  & \\
	$\frac{M^2}{q}\Phi_{r}^R$    & $1.6767 \times10^{-4}$  & $1.67730\times10^{-4}$     & $1.6772834\times10^{-4}$\\
	$\frac{M  }{q}\Phi_{\phi}^R$ & $-5.30424\times10^{-3}$ & $-5.30423170\times10^{-3}$ & \\
    \end{tabular}
    \end{ruledtabular}
    \label{tab:regul-force-circular-orbit}
\end{table}
\endgroup
For a circular orbit, a
calculation in the frequency domain is more efficient, and we expect the
results of~\cite{haas:06, diaz-rivera:04} to be much more accurate than our own results.
This fact is reflected in the number of regularization coefficients we can
reliably extract from the numerical data, before being limited by the accuracy
of the numerical method: the frequency-domain calculation
found usable multipole coefficients up to $\ell = 20$, whereas our data for
$\Phi^{\mathrm{R}}_{(0) \ell}$ is
dominated by noise by the time $\ell$ reaches $16$.
Figure~\ref{fig:multipole-coeffs-circular-orbit-t}
shows this behavior.

\subsection{Accuracy of the numerical method}\label{sec:numerical-accuracy}
Several figures of merit can be used to estimate the accuracy of numerical
values for the self-force. 

An estimate for the truncation error arising from cutting short the summation
in
Eq.~\eqref{eqn:regularize-field} at some $\ell_{\text{max}}$ can be calculated
by considering the behavior of the remaining terms for large $\ell$. 
Detweiler et. al.~\cite{detweiler:03a} showed that the remaining terms scale 
as $\ell^{-4}$ for large $\ell$. They find the functional form of the terms to
be
\begin{align}
    \frac{E \mathcal{P}_{3/2}}{(2\ell-3)(2\ell-1)(2\ell+3)(2\ell+5)}
    \text{,}\label{eqn:Eterm}
\end{align}
where $\mathcal{P}_{3/2} = 36\sqrt2$.
We fit a function of this form to the tail end of a plot of the
multipole coefficients to find the coefficient $E$ in Eq.~\eqref{eqn:Eterm}.
Extrapolating to $\ell \rightarrow \infty$ we find that the truncation error
is
\begin{align}
    \epsilon &= \sum_{\ell = \ell_{\text{max}}}^\infty
                [\text{Eq.~\eqref{eqn:Eterm}}] \\
             &= \frac{12\sqrt2 E \ell_{\text{max}}}
                     {(2\ell_{\text{max}}+3)(2\ell_{\text{max}}+1)
		      (2\ell_{\text{max}}-1)(2\ell_{\text{max}}-3)}
\end{align}
where $\ell_{\text{max}}$ is the value at which we cut the summation short. 		  
For all but
the special case of the $(0)$ component for a circular
orbit, for which all regularization parameters vanish identically, we use
this approach to calculate an estimate for the truncation error.

A second source of error lies in the numerical calculation of the retarded
solution to the wave equation. This error depends on the step size $h$ used to
evolve the field forward in time. For a numerical scheme of a given
convergence order, we can estimate this discretization error by extrapolating
the differences of simulations using different step sizes down to 
$h = 0$. This is what was done in the graphs shown in
Sec.~\ref{sec:convergence-tests-particle}.

We display results for mildly eccentric orbits. A high eccentricity causes 
$\partial_r \Phi$ (displayed in Fig.~\ref{fig:sourced-regression-derivative}) 
to be plagued by high frequency noise produced by
effects similar to those described in
Sec.~\ref{sec:convergence-tests-particle}. This makes it impossible to
reliably estimate the discretization error for these orbits. We do not expect
this to be very different from the errors for mildly eccentric orbits.

Finally we compare our final results for the self-force $F_\alpha$ to
``reference values''. For circular orbits, frequency-domain calculations are
much more accurate than our time-domain computations. We thus compare our results to
the results obtained in~\cite{haas:06}. 
Table~\ref{tab:numerical-errors} lists typical values for the various errors
listed above.
\begin{table}[htbp]
    \centering
    \begin{ruledtabular}
    \begin{tabular}{lc}
	error estimation & mildly eccentric orbit \\
	\hline
	truncation error ($\frac{M^2}{q} \Phi_{(+)}$)
	& $\approx2\times10^{-3}\%$ \\
	discretization error ($\frac{M^2}{q} \partial_r\Phi_{\ell m}$)
	& $\approx10^{-5}\%$ \\
	\hline\hline
	comparison with reference values & circular orbit\\
	\hline
	$\frac{M^2}{q^2} F_t$ 
	& $0.2\%$\\
	$\frac{M^2}{q^2} F_r$ 
	& $0.04\%$\\
	$\frac{M}{q^2} F_\phi$ 
	& $2\times10^{-4}\%$
    \end{tabular}
    \end{ruledtabular}
    \caption{Estimated values for the various errors in the components of the
    self-force as described in the text.
    We show the truncation and discretization errors for a mildly eccentric
    orbit and the total error for a circular orbit. 
    The truncation error is calculated using a plot similar to the one shown
    in 
    Fig.~\ref{fig:zoom-whirl-orbit-multipole-coefficients-zooming}.
    The
    discretization error is estimated using a plot similar to that in
    Fig.~\ref{fig:sourced-regression-derivative} for the $\ell = 2$, $m = 2$
    mode, and the total error is
    estimated as the difference between our values and those
    of~\cite{haas:06}. We use $p = 7.2$ , $e = 0.5$ for the mildly eccentric orbit. 
    Note that we use the tetrad component $\Phi_{(+)}$ for
    the truncation error and the vector component $\partial_r \Phi$ for the
    discretization error. Both are related by the translation table
    Eqs.~\eqref{eqn:translationtable-begin} --
    \eqref{eqn:translationtable-end}, we expect corresponding errors to be
    comparable for $\Phi_{(+)}$ and $\partial_r \Phi$.}
    \label{tab:numerical-errors}
\end{table}

\section{Sample results}\label{sec:sample-results}
In this section we describe some results of our numerical calculation.
\subsection{Mildly eccentric orbit}\label{sec:mildly-eccentric-orbit}
\begin{figure}[tbp]
    \centering
    \includegraphics{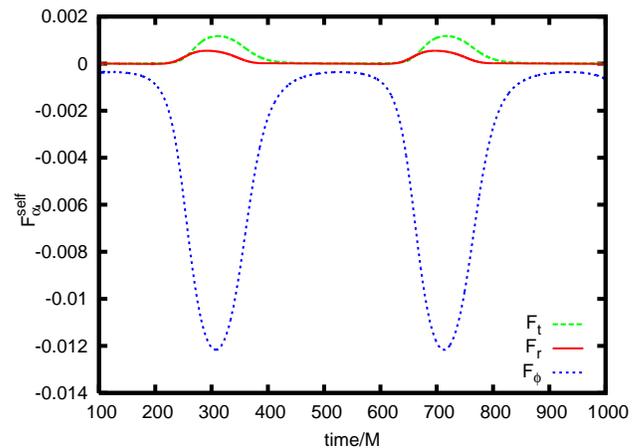}
    \caption{Regularized dimensionless self-force $\frac{M^2}{q^2} F_t$,
    $\frac{M^2}{q^2} F_r$ and $\frac{M}{q^2} F_\phi$
    on a particle on an eccentric orbit with
    $p = 7.2$, $e = 0.5$. }
    \label{fig:regularized-force}
\end{figure}

We choose a particle 
on an eccentric orbit with $p = 7.2$, $e = 0.5$ which
starts at $r = pM/(1-e^2)$, halfway between periastron and apastron.
The field is evolved for $1000\,M$ with a resolution of 16 grid points
per $M$, both in the $t$ and $r^*$ directions, for $\ell=0$. Higher values of
$\ell$ (and thus $m$) require a corresponding increase in the number of grid
points used to achieve the same fractional accuracy. Multipole coefficients
for $0 \le \ell \le 15$ are calculated and used to reconstruct the regularized
self-force $F_\alpha$ along the geodesic. 
Figure~\ref{fig:regularized-force} shows the result of the calculation. 
\begin{figure}[tbp]
    \centering
    \includegraphics[bb=50 50 220 235]{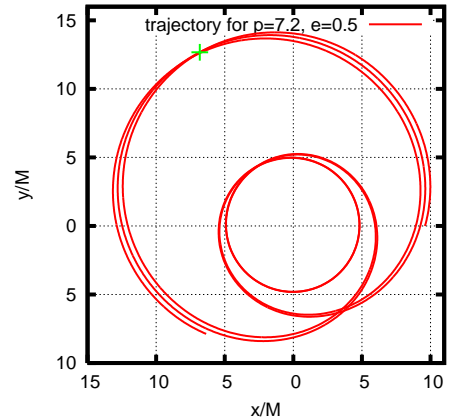}
    \caption{Trajectory of a particle with $p=7.2$, $e=0.5$. The cross-hair
    indicates the point where the data for
    Fig.~\ref{fig:multipole-coeffs-eccentric-orbit-plus} was extracted.}
    \label{fig:eccentric-orbit-trajectory}
\end{figure}
For the choice of parameters used to calculate the force
shown in Fig.~\ref{fig:regularized-force}, the error bars corresponding to the
truncation error (which are already much larger than than the discretization
error)
would be of the order of the line thickness and have not been drawn. 

Already for this small eccentricity, we see that the self-force is most important
when the particle is closest to the black hole (ie.\ for $200\,M \lesssim t
\lesssim 400\,M$ and $600\,M \lesssim t \lesssim 800\,M$); the
self-force acting on the particle is very small once the particle has moved
away to $r \approx 15\,M$. 

\subsection{Zoom-whirl orbit}\label{sec:zoom-whirl-orbit}
\begin{figure}[tbp]
    \centering
    \includegraphics[bb=50 50 220 235]{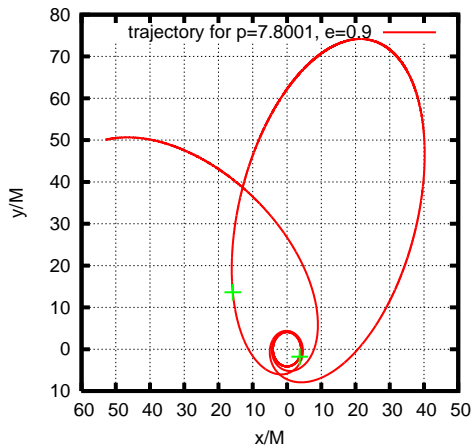}
    \caption{Trajectory of a particle on a zoom-whirl orbit with $p =
    7.8001$, $e=0.9$. The cross-hairs indicate the positions where the data
    shown in Fig.~\ref{fig:zoom-whirl-orbit-multipole-coefficients-zooming}
    and~\ref{fig:zoom-whirl-orbit-multipole-coefficients-whirling} was
    extracted.}
    \label{fig:zoom-whirl-orbit-trajectory}
\end{figure}

\begin{figure}[tbp]
    \centering
    \includegraphics{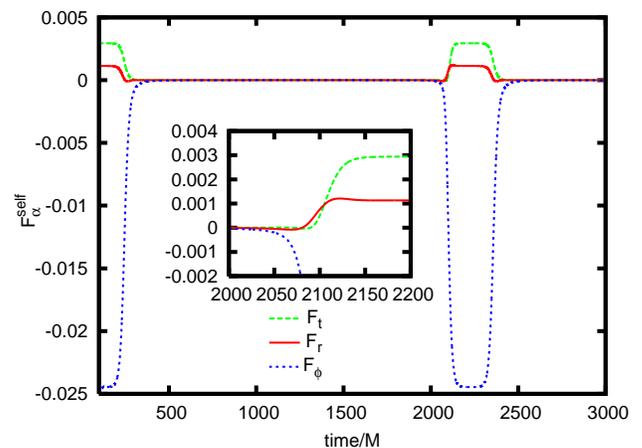}
    \caption{Self-force acting on a particle. Shown is the dimensionless
    self-force $\frac{M^2}{q^2} F_t$,
    $\frac{M^2}{q^2} F_r$ and $\frac{M}{q^2} F_\phi$ on a zoom-whirl
    orbit with $p = 7.8001$, $e=0.9$. The inset shows a magnified view of the
    self-force when the particle is about to enter the whirl phase. 
    No error bars showing an estimate error are shown, as the errors shown eg.
    in Table~\ref{tab:numerical-errors} are to small to show up on the graph.
    Notice that
    the self-force is essentially zero during the zoom phase $500\,M \lesssim
    t \lesssim 2000\,M$ and reaches a constant value very quickly after the
    particle enters into the whirl phase.}
    \label{fig:zoom-whirl-orbit-force}
\end{figure}

Highly eccentric orbits are of most interest as sources of gravitational
radiation. For nearly parabolic orbits with $e \lesssim 1$ and $p \gtrsim 6 +
2e$, a particle revolves around the black hole a number of times,
moving on a nearly circular trajectory close to the event horizon (``whirl
phase''), before 
moving away from the black hole (``zoom phase''). During the whirl phase
the particle is in the strong field region of the black
hole, emitting copious amounts of radiation.
Figures~\ref{fig:zoom-whirl-orbit-trajectory} 
and~\ref{fig:zoom-whirl-orbit-force} show
the trajectory of a particle and the force 
on such an orbit with $p = 7.8001$, $e = 0.9$.
Even more so than for the mildly eccentric orbit discussed in
Sec.~\ref{sec:mildly-eccentric-orbit}, the self-force (and thus the amount
of radiation produced) is much larger while the particle is close to the black
hole than when it zooms out. 

Defining energy $E$ per unit mass and
angular momentum $L$ per unit mass in the usual way,
\begin{align}
    E &= - \left(\diff{}{t}\right)^\alpha u_\alpha\text{,} & 
    L &=   \left(\diff{}{\phi}\right)^\alpha u_\alpha\text{,}
\end{align}
and following eg.\ the treatment of Wald~\cite{wald:84}, Appendix~C, it is easy
to see that the rates of change $\dot E$ and $\dot L$ (per unit proper time)
are
directly related to components of the acceleration $a_\alpha$ 
(and therefore force) experienced by the particle via
\begin{align}
    \dot E &= -a_t\text{,} & 
    \dot L &=  a_\phi\text{.}
\end{align}
The self-force shown in Fig.~\ref{fig:zoom-whirl-orbit-force}
therefore confirms our na\"\i{}ve expectation that the self-force should
decrease both the energy and angular momentum of the particle as radiation is
emitted. 

\begin{figure}[tbp]
    \centering
    \includegraphics{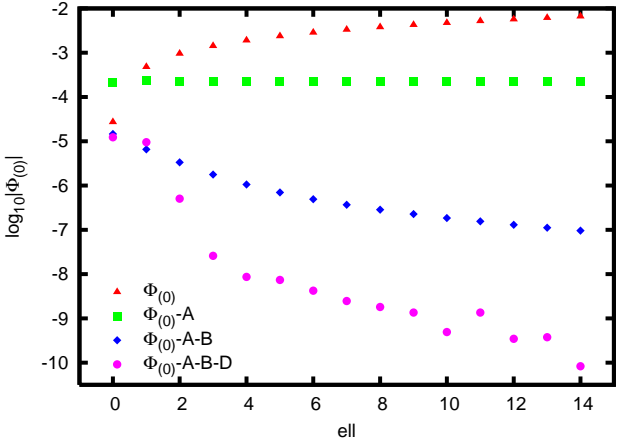}
    \caption{Multipole coefficients of $\frac{M^2}{q}\real\Phi^{\mathrm{R}}_{(0)}$ for a
    particle on a zoom-whirl orbit ($p = 7.8001$, $e = 0.9$).
    The
    coefficients are extracted at $t = 2000\,M$ as the particle is about to
    enter the whirl phase. As $\dot r$ is non-zero, all components of the
    self-force require regularization and we see that the dependence of the
    multipole coefficients on $\ell$ is as predicted by
    Eq.~\ref{eqn:singular-field}. After the removal of the regularization
    parameters $A_{(\mu)}$, $B_{(\mu)}$, and $D_{(\mu)}$ the remainder is
    proportional to $\ell^0$, $\ell^{-2}$ and $\ell^{-4}$ respectively.}
    \label{fig:zoom-whirl-orbit-multipole-coefficients-zooming}
\end{figure}

\begin{figure}[tbp]
    \centering
    \includegraphics{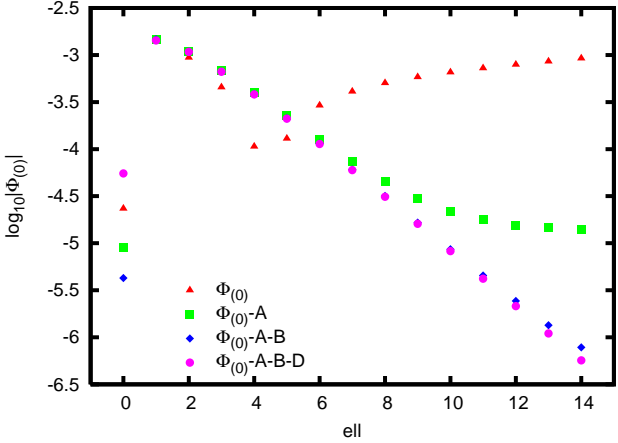}
    \caption{Multipole coefficients of $\real\Phi^{\mathrm{R}}_{(0)}$ for a
    particle on a zoom-whirl orbit ($p = 7.8001$, $e = 0.9$).
    The
    coefficients are extracted at $t = 2150\,M$ while the particle is in the
    whirl phase. The orbit is nearly circular at this time, 
    causing the dependence on $\ell$ after removal of the
    regularization parameters to approximate that of a true circular orbit.}
    \label{fig:zoom-whirl-orbit-multipole-coefficients-whirling}
\end{figure}

It is instructive to have a closer look at the force acting on the particle
when it is within the zoom phase, and also when it is moving 
around the
black hole on the nearly circular orbit of the whirl phase. In
Fig.~\ref{fig:zoom-whirl-orbit-multipole-coefficients-zooming} and
Fig.~\ref{fig:zoom-whirl-orbit-multipole-coefficients-whirling} we show
plots of $\Phi_{(0)\ell}$ vs.\ $\ell$ after the removal of the  
$A_{(\mu)}$, $B_{(\mu)}$, and $D_{(\mu)}$ terms. While the particle is still
zooming in toward the black hole, $\Phi_{(0)\ell}$ behaves exactly as for the
mildly eccentric orbit described in Sec.~\ref{sec:mildly-eccentric-orbit}
over the full range of $\ell$ plotted; ie.\
the magnitude of each term scales as $\ell^0$, $\ell^{-2}$ and $\ell^{-4}$, after
removal of the $A_{(\mu)}$, $B_{(\mu)}$, and $D_{(\mu)}$ terms respectively.
Close to the black hole, on the
other hand, the particle moves along a nearly circular trajectory. If the orbit
were perfectly circular for all times, ie.\ $\dot r \equiv 0$, then the $(0)$ component
would not require regularization at all, and the multipole coefficients would decay
exponentially, resulting in a straight line on the semi-logarithmic plot shown
in Fig.~\ref{fig:zoom-whirl-orbit-multipole-coefficients-whirling}. As the
real orbit is not precisely circular, curves eventually deviate from a
straight line.
Removal of the $A_{(\mu)}$
term is required almost immediately (beginning with $\ell \approx 3$), while
the $D_{(\mu)}$ term starts to become important only after $\ell \approx 11$. 
This shows that there is a smooth transition from the self-force on a
circular orbit, which does not require regularization for the $t$ and $\phi$
components, to that of a generic orbit, for which all components of the self-force
require regularization.

\begin{acknowledgments}
    We thank Eric Poisson and Eran Rosenthal for useful discussions and
    suggestions. This work was supported by the Natural Sciences and
    Engineering Council of Canada.
\end{acknowledgments}

\appendix 
\section{Translation tables}\label{sec:translation-tables}
We quote the results of~\cite{haas:06} for the translation table between the
modes $\Phi_{\ell m}$ and the tetrad components $\Phi_{(\mu) \ell m}$ with
respect to the pseudo-Cartesian basis
\begin{align} 
    e^\alpha_{(0)} &= \biggl[ \frac{1}{\sqrt{f}}, 0, 0, 0 \biggr]\text{,} \\ 
    e^\alpha_{(1)} &= \biggl[ 0, \sqrt{f}\sin\theta\cos\phi, 
    \frac{1}{r} \cos\theta\cos\phi, -\frac{\sin\phi}{r\sin\theta}
    \biggr]\text{,}\\ 
    e^\alpha_{(2)} &= \biggl[ 0, \sqrt{f}\sin\theta\sin\phi, 
    \frac{1}{r} \cos\theta\sin\phi, \frac{\cos\phi}{r\sin\theta} 
    \biggr]\text{,}\\ 
    e^\alpha_{(3)} &= \biggl[ 0, \sqrt{f}\cos\theta, 
    -\frac{1}{r} \sin\theta, 0 \biggr]\text{,} 
\end{align} 
and the complex combinations
$e^\alpha_{(\pm)} \define e^\alpha_{(1)} \pm i e^\alpha_{(2)}$,
\begin{align} 
e^\alpha_{(\pm)} = \biggl[ 0, \sqrt{f}\sin\theta e^{\pm i \phi},   
\frac{1}{r} \cos\theta e^{\pm i\phi}, 
\frac{\pm i e^{\pm i\phi}}{r\sin\theta} \biggr]\text{.}  
\end{align}
With these, the spherical-harmonic modes
$\Phi_{(\mu) \ell m}(t,r)$ are given in terms of $\Phi_{\ell m}(t,r)$ by  
\begin{align}
    \Phi_{(0)\ell m} =& \frac{1}{\sqrt{f}} 
    \frac{\partial}{\partial t} \Phi_{\ell m}\text{,} 
    \label{eqn:translationtable-begin}
    \\ 
    \Phi_{(+)\ell m} =& -\sqrt{ \frac{(\ell+m-1)(\ell+m)}{(2\ell-1)(2\ell+1)} }  
    \nonumber\\\mbox{}&\times
    \left( \sqrt{f} \frac{\partial}{\partial r} 
	   - \frac{\ell-1}{r} \right) \Phi_{\ell-1,m-1} 
    \nonumber \\ \mbox{}&
    + \sqrt{ \frac{(\ell-m+1)(\ell-m+2)}{(2\ell+1)(2\ell+3)} }  
    \nonumber\\\mbox{}&\times
    \left( \sqrt{f} \frac{\partial}{\partial r} 
    + \frac{\ell+2}{r} \right) \Phi_{\ell+1,m-1}\text{,} 
    \\ 
    \Phi_{(-)\ell m} =& \sqrt{ \frac{(\ell-m-1)(\ell-m)}{(2\ell-1)(2\ell+1)} }  
    \nonumber\\\mbox{}&\times
    \left( \sqrt{f} \frac{\partial}{\partial r} 
	   - \frac{\ell-1}{r} \right) \Phi_{\ell-1,m+1} 
    \nonumber \\ \mbox{}&
    - \sqrt{ \frac{(\ell+m+1)(\ell+m+2)}{(2\ell+1)(2\ell+3)} }
    \nonumber\\\mbox{}&\times
    \left( \sqrt{f} \frac{\partial}{\partial r} 
    + \frac{\ell+2}{r} \right) \Phi_{\ell+1,m+1}\text{,}
    \\ 
    \Phi_{(3)\ell m} =& \sqrt{ \frac{(\ell-m)(\ell+m)}{(2\ell-1)(2\ell+1)} }  
    \left( \sqrt{f} \frac{\partial}{\partial r} 
	   - \frac{\ell-1}{r} \right) \Phi_{\ell-1,m} 
    \nonumber \\   \mbox{} &
    + \sqrt{ \frac{(\ell-m+1)(\ell+m+1)}{(2\ell+1)(2\ell+3)} }  
    \nonumber\\\mbox{}&\times
    \left( \sqrt{f} \frac{\partial}{\partial r} 
    + \frac{\ell+2}{r} \right) \Phi_{\ell+1,m}\text{.}
    \label{eqn:translationtable-end}
\end{align}

\section{Regularization parameters}\label{sec:regularization-parameters}
For completeness we list the regularization parameters as calculated
in~\cite{haas:06}. Quantities
bearing a subscript ``$0$'' are evaluated at the particle's position.
\begin{align} 
A_{(0)} &= \frac{\dot r_0}{\sqrt{f_0}(r_0^2+L^2)} \operatorname{sign}(\Delta)\text{,}  \\ 
A_{(+)} &= -e^{i\phi_0} \frac{E}{\sqrt{f_0}(r_0^2+L^2)} \operatorname{sign}(\Delta)\text{,}  \\ 
A_{(3)} &= 0\text{,}  
\end{align} 
where $f_0 \define 1 - 2M/r_0$ and $\operatorname{sign}(\Delta)$ is equal to $+1$ if
$\Delta > 0$ and to $-1$ if $\Delta < 0$. We have, in addition,  
$A_{(-)} = \bar{A}_{(+)}$, $A_{(1)} = \real[A_{(+)}]$, and 
$A_{(2)} = \imag[A_{(+)}]$.  

We also use 
\begin{align} 
B_{(0)} &= -\frac{E r_0 \dot r_0}{\sqrt{f_0}(r_0^2+L^2)^{3/2}} \mathcal{E}  
+ \frac{E r_0 \dot r_0}{2\sqrt{f_0}(r_0^2+L^2)^{3/2}} \mathcal{K}\text{,}  \\ 
B_{(+)} &= e^{i\phi_0} \bigl( B^{c}_{(+)} - i B^{s}_{(+)}
\bigr)\text{,}  \\ 
B^{c}_{(+)} &= \biggl[ 
\frac{r_0 \dot r_0^2}{\sqrt{f_0}(r_0^2+L^2)^{3/2}} 
+ \frac{\sqrt{f_0}}{2 r_0 \sqrt{r_0^2+L^2}} 
\biggr] \mathcal{E} 
\nonumber \\ &  \mbox{} 
- \biggl[ 
\frac{r_0 \dot r_0^2}{2\sqrt{f_0}(r_0^2+L^2)^{3/2}} 
+ \frac{\sqrt{f_0}-1}{r_0 \sqrt{r_0^2+L^2}} 
\biggr] \mathcal{K}\text{,} \\ 
B^{s}_{(+)} &=
- \frac{(2-\sqrt{f_0})\dot r_0}{2L\sqrt{r_0^2+L^2}\sqrt{f_0}} \mathcal{E}  
+ \frac{(2-\sqrt{f_0})\dot r_0}{2L\sqrt{r_0^2+L^2}\sqrt{f_0}} \mathcal{K}\text{,} \\ 
B_{(3)} &= 0\text{.} 
\end{align} 
In addition, $B_{(-)} = \bar{B}_{(+)}$, $B_{(1)} =
\real[B_{(+)}] = B^{c}_{(+)} \cos\phi_0 + B^{s}_{(+)} 
\sin\phi_0$, and $B_{(2)} = \imag[B_{(+)}] = B^{c}_{(+)}
\sin\phi_0 - B^{s}_{(+)} \cos\phi_0$. 

Here, the rescaled elliptic integrals $\mathcal{E}$ and $\mathcal{K}$ are
defined by
\begin{align} 
\mathcal{E} \define \frac{2}{\pi} \int_0^{\pi/2} (1-k\sin^2\psi)^{1/2}\, \d\psi  
= F\Bigl(-{\frac{1}{2}},{\frac{1}{2}}; 1; k\Bigr)
\end{align} 
and 
\begin{align} 
\mathcal{K} \define \frac{2}{\pi} \int_0^{\pi/2} (1-k\sin^2\psi)^{-1/2}\, \d\psi     
= F\Bigl({\frac{1}{2}},{\frac{1}{2}}; 1; k\Bigr)\text{,} 
\end{align} 
in which $k \define L^2/(r_0^2 + L^2)$. 

We also use 
\begin{align}
C_{(\mu)} = 0
\end{align} 
and 
\begin{widetext}
\begin{align} 
D_{(0)} &= -\biggl[ 
\frac{E r_0^3 (r_0^2-L^2) \dot r_0^3}
{2\sqrt{f_0}(r_0^2+L^2)^{7/2}} 
+ \frac{E(r_0^7 + 30M r_0^6 - 7 L^2 r_0^5 + 114 M L^2 r_0^4 
+ 104 M L^4 r_0^2 + 36 M L^6) \dot r_0}
  {16 r_0^4 \sqrt{f_0}(r_0^2+L^2)^{5/2}} 
\biggr] \mathcal{E} 
\nonumber \\ &  \mbox{} 
+ \biggl[ 
\frac{E r_0^3 (5 r_0^2 - 3 L^2) \dot r_0^3}
  {16\sqrt{f_0}(r_0^2+L^2)^{7/2}} 
+ \frac{E(r_0^5 + 16M r_0^4 - 3 L^2 r_0^3 + 42 M L^2 r_0^2 
+ 18 M L^4) \dot r_0}{16 r_0^2
  \sqrt{f_0}(r_0^2+L^2)^{5/2}} \biggr] \mathcal{K}\text{,} \\ 
D_{(+)} &= e^{i\phi_0} \bigl( D^{c}_{(+)} - i D^{s}_{(+)} 
\bigr)\text{,}  \\ 
D^{c}_{(+)} &= 
\biggl[ \frac{r_0^3 (r_0^2 - L^2) \dot r_0^4}
{2\sqrt{f_0}(r_0^2+L^2)^{7/2}} 
- \frac{r_0 \dot r_0^2}{4(r_0^2+L^2)^{3/2}} 
+ \frac{(3r_0^7 + 6 M r_0^6 - L^2 r_0^5 + 31 M L^2 r_0^4 
  + 26 M L^4 r_0^2 + 9 M L^6) \dot r_0^2}
  {4 r_0^4 \sqrt{f_0} (r_0^2+L^2)^{5/2}} 
\nonumber \\ &  \mbox{} 
+ \frac{(3r_0^7 + 8 M r_0^6 + L^2 r_0^5 + 26 M L^2 r_0^4 
  + 22 M L^4 r_0^2 + 8 M L^6) \sqrt{f_0}}
  {16 r_0^6 (r_0^2+L^2)^{3/2}} 
- \frac{r_0^3 + 2 M r_0^2 + 4 M L^2}
  {8 r_0^4 \sqrt{r_0^2+L^2}} \bigg] \mathcal{E} 
\nonumber \\ &  \mbox{} 
+ \biggl[ -\frac{r_0^3 (5 r_0^2 - 3 L^2) \dot r_0^4}
 {16\sqrt{f_0}(r_0^2+L^2)^{7/2}} 
+ \frac{r_0 \dot r_0^2}{8(r_0^2+L^2)^{3/2}} 
- \frac{(7 r_0^5 + 12 M r_0^4 - L^2 r_0^3 
  + 46 M L^2 r_0^2 + 18 M L^4) \dot r_0^2}
  {16 r_0^2 \sqrt{f_0} (r_0^2+L^2)^{5/2}} 
\nonumber \\ &  \mbox{} 
- \frac{(7r_0^5 + 6 M r_0^4 + 6 L^2 r_0^3 + 12 M L^2 r_0^2 
  + 4 M L^4) \sqrt{f_0}}
  {16 r_0^4 (r_0^2+L^2)^{3/2}} 
+ \frac{3}{8 r_0 \sqrt{r_0^2+L^2}} \biggr] \mathcal{K}\text{,}  \\ 
D^{s}_{(+)} &= 
\biggl[ 
\frac{r_0^2(r_0^2 - 7L^2)(\sqrt{f_0} - 2)\dot r_0^3}
  {16 L \sqrt{f_0} (r_0^2+L^2)^{5/2}} 
- \frac{(2r_0^7 + M r_0^6 + 5 L^2 r_0^5 + 10 M L^2 r_0^4 
  + 29 M L^4 r_0^2 + 14 M L^6) \dot r_0} 
  {8 r_0^5 L (r_0^2+L^2)^{3/2}}
\nonumber \\ &  \mbox{} 
+ \frac{(r_0^5 - M r_0^4 + 4 L^2 r_0^3 - 5 M L^2 r_0^2 
  + 2 M L^4) \dot r_0}
  {4 r_0^3 L \sqrt{f_0}(r_0^2+L^2)^{3/2}} \biggr] \mathcal{E} 
\nonumber \\ &  \mbox{} 
+ \biggl[ 
- \frac{r_0^2(r_0^2 - 3L^2)(\sqrt{f_0} - 2)\dot r_0^3}
  {16 L \sqrt{f_0} (r_0^2+L^2)^{5/2}} 
+ \frac{(4r_0^5 + 2 M r_0^4 + 7 L^2 r_0^3 + 10 M L^2 r_0^2 
  + 14 M L^4) \dot r_0} 
  {16 r_0^3 L (r_0^2+L^2)^{3/2}}
\nonumber \\ &  \mbox{} 
- \frac{(2 r_0^3 - 2 M r_0^2 + 5 L^2 r_0 - 8 M L^2) \dot r_0}
  {8 r_0 L \sqrt{f_0}(r_0^2+L^2)^{3/2}} \biggr] \mathcal{K}\text{,} \\ 
D_{(3)} &= 0\text{.} 
\end{align}
\end{widetext}
And finally, $D_{(-)} = \bar{D}_{(+)}$, $D_{(1)} =
\real[D_{(+)}] = D^{c}_{(+)} \cos\phi_0 + D^{s}_{(+)}
\sin\phi_0$, and $D_{(2)} = \imag[D_{(+)}] = D^{c}_{(+)}
\sin\phi_0 - D^{s}_{(+)} \cos\phi_0$. 
\section{Piecewise polynomials}\label{sec:piecewise-polynomials}
In two places in the numerical simulation we introduce \emph{piecewise
polynomials} to approximate the scalar field $\psi_{\ell m}$ across the
world line, where it is continuous but not
differentiable. By a piecewise polynomial we mean a polynomial of the form
\begin{align}
    p(t, r^*) &= 
    \begin{cases}
	\displaystyle \sum_{n,m = 0}^{N} \frac{c_{nm}}{n!m!} u^n v^m 
	    & \text{if $r^*(u,v) > r_0^*$}\\
	\displaystyle \sum_{n,m = 0}^{N} \frac{c'_{nm}}{n!m!} u^n v^m 
	    & \text{if $r^*(u,v) < r_0^*$}
    \end{cases}
    \text{,}\label{eqn:piecewise-polynomial}
\end{align}
where $u = t-r^*$, $v = t+r^*$ are characteristic coordinates, $r_0^*$ is
the position of the particle at the time $t(u,v)$, and $N$ is the order of the
polynomial, which for our purposes is $N=4$ or less. 
The two sets of coefficients $c_{nm}$ and $c'_{nm}$ are
not independent of each other, but are linked via jump conditions that can be
derived from the wave equation
[Eq.~\eqref{eqn:reduced-wave-eqn}]. To do so, we rewrite the wave equation in
the characteristic coordinates $u$ and $v$ and reintroduce the integral over
the world line on the right-hand side,
\begin{align}
    -4\partial_u\partial_v \psi - V\psi = \int_\gamma \widehat S(\tau)
    \Dirac(u-u_p) \Dirac(v-v_p) \, \d\tau\text{,}
    \label{eqn:uv-wave-eqn}
\end{align}
where 
$\widehat S(\tau) = -8\pi q 
\frac{\conj{Y}_{\ell m}\bm(\pi/2, \phi_p(\tau)\bm)}{r_p(\tau)}$
is the source term and quantities bearing a subscript $p$ are evaluated on the
world line at proper time $\tau$. 

\begin{figure}[tbp]
    \centering
    \includegraphics{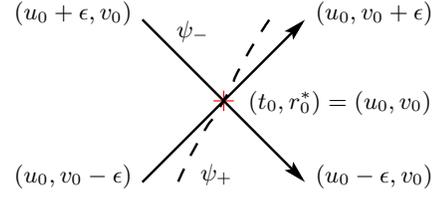}
    \caption{Paths taken in the calculation of the jump conditions.
    $(u_0,v_0)$ denotes an arbitrary but fixed point along the world line
    $\gamma$. The wave equation is integrated along the lines of constant $u$
    or $v$ indicated in the sketch. Note that in order to move from the domain
    on the left to the domain on the right, $u$ has to run from $u_0 +
    \epsilon$ to $u_0 - \epsilon$. Where appropriate we label quantities
    connected to the domain on the left by a subscript ``$-$'' and quantities
    connected to the domain on the right by ``$+$''.}
    \label{fig:crossing-the-world-line}
\end{figure}

Here and in the
following we use the notation
\begin{align}
    \jump{\partial_u^n\partial_v^m\psi} &=
    \lim_{\epsilon \approaches 0^+} 
	[\partial_u^n\partial_v^m\psi(t_0,r_0^*+\epsilon) 
	- \partial_u^n\partial_v^m\psi(t_0, r_0^*-\epsilon)]
\end{align}
to denote the jump in $\partial_u^n\partial_v^m\psi$ across the world line.
First, we notice that the source term does not contain any derivatives of the
Dirac $\Dirac$-function, causing the solution $\psi$ to be continuous. This means
that the zeroth-order jump vanishes: $\jump{\psi} = 0$. 
Our task is then to find the remaining jump conditions at a point $(t_0, r_0^*)$ 
for $n, m \le 4$. Alternatively, instead of crossing the world line
along a line $t = t_0 = \const$ we can also choose to cross along lines of
$u = u_0 = \const$ or $v = v_0 = \const$, noting that for a line of constant
$v$ the coordinate $u$ runs from $u_0 + \epsilon$ to $u_0 - \epsilon$ to cross 
from the left to
the right of the world line. Figure~\ref{fig:crossing-the-world-line} provides a
clearer description of the paths taken.

In order to find the jump $\jump{\partial_u \psi}$ we integrate the wave
equation along the line $u=u_0$ from $v_0-\epsilon$ to $v_0+\epsilon$
\begin{align}
    -4\int_{v_0-\epsilon}^{v_0+\epsilon}&\partial_u\partial_v\psi \d v
    - \int_{v_0-\epsilon}^{v_0+\epsilon} V \psi \d v =
    \nonumber\\\mbox{}&
    \int_\gamma \widehat S(\tau) \Dirac(u_0-u_p) 
	\int_{v_0-\epsilon}^{v_0+\epsilon} \Dirac(v-v_p) \d v \, \d\tau
	\text{,}
\end{align}
which, after involving 
$\int_{v_0-\epsilon}^{v_0+\epsilon} \Dirac(v-v_p) \d v =
\Heaviside(v_p-v_0+\epsilon) \Heaviside(v_0-v_p+\epsilon)$ and 
$\Dirac\bm(g(x)\bm)=\Dirac(x-x_0)/\abs{g'(x_0)}$, yields
\begin{align}
    \jump{\partial_u \psi} &= -\frac14 \frac{f_0}{E - \dot r_0}
    \widehat S(\tau_0)\text{,}
\end{align}
where the overdot denotes differentiation with respect to proper time $\tau$.

Similarly, after first taking a derivative of the wave equation with respect to
$v$ and integrating from $u_0 + \epsilon$ to $u_0 - \epsilon$, we obtain
\begin{align}
    -4\int_{u_0+\epsilon}^{u_0-\epsilon}&\partial_u\partial^2_v\psi \d u
    - \int_{u_0+\epsilon}^{u_0-\epsilon} V \psi \d u =
    \nonumber\\\mbox{}&
    \int_\gamma \widehat S(\tau) 
    \int_{u_0+\epsilon}^{u_0-\epsilon} \Dirac(u-u_p) \d u \,
    \Dirac'(v_0-v_p) \d\tau \text{.}
\end{align}
We find
\begin{align}
    \jump{\partial^2_v \psi} &=
    \frac14\frac{f_0}{E+\dot r_0} \odiff{}{\tau}
    \Bigl[\frac{f_p}{E+\dot r_p} \widehat
    S(\tau)\Bigr]_{\left|\tau = \tau_0\right.}\text{.}
\end{align}

Systematically repeating this procedure we find expressions for the
jumps in all
the derivatives that are purely in the $u$ or $v$ direction.
Table~\ref{tab:pure-jumps} lists these results.
\begin{table}
    \begin{ruledtabular}
    \begin{align*}
	\jump{\psi} =& 0
	\\
	\jump{\partial_u \psi} =& -\frac14 \mdef_0^{-1} \widehat S(\tau_0) 
	\text{, }
	\jump{\partial_v \psi} =  \frac14 \pdef_0^{-1} \widehat S(\tau_0)
	\\
	\jump{\partial^2_u \psi} =& 
	-\frac14 \mdef_0^{-1} \odiff{ }{\tau} \Bigl( \mdef_p^{-1}
	  \widehat S(\tau) \Bigr)_{\left|\tau = \tau_0\right.}
	\\
	\jump{\partial^2_v \psi} =&  
	\frac14 \pdef_0^{-1} 
	  \odiff{}{\tau} \Bigl( \pdef_p^{-1}
	  \widehat S(\tau) \Bigr)_{\left|\tau = \tau_0\right.}
	\\
	\jump{\partial^3_u \psi} =& 
	 \frac14 V \pdef_0 \mdef_0^{-1} \jump{\partial_u \psi}
	  - \frac14 \mdef_0^{-1} 
	  \odiff{}{\tau} \Bigl[\mdef_p^{-1}
	  \odiff{}{\tau} \Bigl(\mdef_p^{-1} 
	  \widehat S(\tau) \Bigr)\Bigr]_{\left|\tau = \tau_0\right.}
	\\
	\jump{\partial^3_v \psi} =&
	  \frac14 V \mdef_0 \pdef_0^{-1} \jump{\partial_v \psi}
	  + \frac14 \pdef_0^{-1} 
	  \odiff{ }{\tau} \Bigl[\pdef_p^{-1} 
	  \odiff{}{\tau} \Bigl(\pdef_p^{-1} 
	  \widehat S(\tau) \Bigr)\Bigr]_{\left|\tau = \tau_0\right.}
	\\
	\jump{\partial^4_u \psi} =&
	-\frac14 \Bigl[
	  -\frac12 \mdef_0^{-1} V \frac{\ddot r_0}{E}
	  +\frac12 \mdef_0^{-1} V \odiff{}{\tau} 
	    \Bigl(\frac{f_p}{E} \pdef_p^2 \mdef_p^{-1}\Bigr)
	    _{\left|\tau = \tau_0\right.}
	  \\\mbox{}&
	  +3 \pdef_0 \mdef_0^{-1} \partial_u V
	  +\pdef_0^2 \mdef_0^{-2} \partial_v V
	\Bigr] \jump{\partial_u \psi}
	+\frac12 \pdef_0 \mdef_0^{-1} V \jump{\partial^2_u \psi}
	\\\mbox{}&
	-\frac14 \mdef_0^{-1}\odiff{}{\tau}\bm{\Bigl(}
	  \mdef_p^{-1}\odiff{}{\tau}\Bigl\{
	  \mdef_p^{-1}\odiff{}{\tau}\Bigl[
	  \mdef_p^{-1} \widehat S(\tau)
	  \Bigr]\Bigr\}\bm{\Bigr)}_{\left|\tau = \tau_0\right.}
	\\
	\jump{\partial^4_v \psi} =&
	\frac14 \Bigl[
	  -\frac12 \pdef_0^{-1} V \frac{\ddot r_0}{E}
	  +\frac12 \pdef_0^{-1} V \odiff{}{\tau} 
	    \Bigl(\frac{f_p}{E} \mdef_p^2 \pdef_p^{-1}\Bigr)
	    _{\left|\tau = \tau_0\right.}
	  \\\mbox{}&
	  +3 \mdef_0 \pdef_0^{-1} \partial_v V
	  +\mdef_0^2 \pdef_0^{-2} \partial_u V
	\Bigr] \jump{\partial_v \psi}
	-\frac12 \mdef_0 \pdef_0^{-1} V \jump{\partial^2_v \psi}
	\\\mbox{}&
	-\frac14 \pdef_0^{-1}\odiff{}{\tau}\bm{\Bigl(}
	  \pdef_p^{-1}\odiff{}{\tau}\Bigl\{
	  \pdef_p^{-1}\odiff{}{\tau}\Bigl[
	  \pdef_p^{-1} \widehat S(\tau)
	  \Bigr]\Bigr\}\bm{\Bigr)}_{\left|\tau = \tau_0\right.}
    \end{align*}
    \end{ruledtabular}
    \caption{Jump conditions for the derivatives purely in the $u$ or $v$
    directions. $\dot r$ and $\ddot r$ are the particle's radial velocity and
    acceleration, respectively. They are obtained from the equation of motion
    for the particle. $\mdef \define \frac{E-\dot r}{f}$ and 
    $\pdef \define \frac{E + \dot r}{f}$ were introduced for notational
    convenience.
    Quantities bearing a subscript $p$ are evaluated on the
    particle's world line, while quantities bearing a subscript $0$ are
    evaluated at the particle's current position. Derivatives of $V$ with
    respect to either $u$ or $v$ are evaluated as $\partial_u V = -\frac12 f
    \partial_r V$ and $\partial_v V = \frac12 f \partial_r V$, respectively.}
    \label{tab:pure-jumps}
\end{table}
Jump conditions for derivatives involving both $u$ and $v$ are obtained
directly from the wave equation
[Eq.~\eqref{eqn:uv-wave-eqn}]. We see that
\begin{align}
    \jump{\partial_u \partial_v \psi} &= 0\text{,}
\end{align}
and taking an additional derivative with respect to $u$ on both sides reveals
that
\begin{align}
    \jump{\partial_u^2 \partial_v \psi} &= -\frac14 V \jump{\partial_u \psi}
    \text{.}
\end{align}
Systematically repeating this procedure we can find jump conditions for each
of the mixed derivatives by evaluating
\begin{align}
    \jump{\partial^{n+1}_u \partial^{m+1}_v \psi} &=
      -\frac14 \jump{\partial^n_u \partial^m_v (V \psi)}
      \text{,}\label{eqn:mixed-jumps}
\end{align}
where $n, m \ge 0$ and derivatives of $V$ with respect to either $u$ or $v$
are evaluated as $\partial_u V = -\frac12 f \partial_r V$ and $\partial_v V =
\frac12 f \partial_r V$, respectively.

The results of Table~\ref{tab:pure-jumps} and~Eq.~\eqref{eqn:mixed-jumps}
allow us to express the coefficients of the left-hand polynomial in
Eq.~\eqref{eqn:piecewise-polynomial} in terms of the jump conditions and the
coefficients of the right-hand side:
\begin{align}
    c'_{nm} &= c_{nm} - \jump{\partial_u^n \partial_v^m \psi}\text{.}
\end{align}

For $N=4$ this leaves us with $25$ unknown coefficients $c_{nm}$ which can be
uniquely determined by demanding that the polynomial match the value of the 
field on the 25 grid points surrounding the particle.
When we are interested in integrating the polynomial, as in the case
of the potential term in the fourth-order algorithm, we do not need all these terms.
Instead, in order to calculate e.g. the integral ${\int\!\!\int}_{\text{cell}} V
\psi \, \d u \, \d v$ up to terms of order $h^5$, as is needed to achieve
overall $\orderof{h^4}$ convergence, it is sufficient to
include only terms such that $n+m \le 2$, thus reducing the number of unknown
coefficients to $6$. In this case Eq.~\eqref{eqn:piecewise-polynomial} becomes
\begin{align}
    p(t, r^*) &= 
    \begin{cases}
	\displaystyle \sum_{m+n \le 2} \frac{c_{nm}}{n!m!} u^n v^m 
	    & \text{if $r^*(u,v) > r_0^*$}\\
	\displaystyle \sum_{m+n \le 2} \frac{c'_{nm}}{n!m!} u^n v^m 
	    & \text{if $r^*(u,v) < r_0^*$}
    \end{cases}
    \text{.}
\end{align}
The six coefficients can then be determined by matching the polynomial
to the field values at the six grid points 
which 
lie within the past light cone of the
grid point whose field value we want to calculate. 

\bibliography{paper}
\end{document}